\documentclass[traditabstract]{aa} % for the abstract without structuration

\usepackage{txfonts}
\usepackage[T1]{fontenc}
\usepackage{graphicx}
\usepackage{txfonts}
\def\la{\raise.5ex\hbox{$<$}\kern-.8em\lower 1mm\hbox{$\sim$}}
\def\ma{\raise.5ex\hbox{$>$}\kern-.8em\lower 1mm\hbox{$\sim$}}

\def\msol{M$_{\odot}$ }

\def\kms{$\rm km\, s^{-1}$}
\def\cm3{$\rm cm^{-3}$}
\def\Ts{$\rm T_{*}$~}
\def\Vs{$\rm V_{s}$}
\def\n0{$\rm n_{0}$}
\def\B0{$\rm B_{0}$}

\def\Ne{$\rm N_{e}$~}
\def\Te{$\rm T_{e}$~}

\def\erg{$\rm erg\, cm^{-2}\, s^{-1}$}

\def\L12{L$_{12\mu m}$~}
\def\F12{F$_{12\mu m}$~}

\def\Hb{H${\beta}$}
\def\Ha{H${\alpha}$}
\def\Hg{H${\gamma}$}

\def\RO3{R$_{[OIII]}$}

\begin{document}

   \title{
  Analysis of metal-poor galaxy spectra  in the redshift range 0.00574-0.05368}

   \author{M. Contini \inst{1,2}
}

   \institute{Dipartimento di Fisica e Astronomia "G.Galilei", University of Padova, Vicolo dell'Osservatorio 3. 35122 Padova, Italy
         \and
             School of Physics and Astronomy, Tel Aviv University, Tel Aviv 69978, Israel\\
%             \email{contini@tauex.tau.ac.il}
}

   \date{Received }

% \abstract{}{}{}{}{}
% 5 {} token are mandatory

  \abstract{
We  present  an analysis of the metal-poor galaxy spectra in the redshift range 0.00574$\leq$z$\leq$0.05368
which were  reported by Nakajima et al (2022) in their  EMPG (extreme metal poor galaxy) sample. 
The models account for the  active galactic nuclei (AGN) and the starburst (SB) galaxies, for accretion and ejection, 
for the physical parameters and the element abundances.
The results  are obtained in particular for the two cases,
 the emitting nebula is  ejected outward from the galaxy radiation source (RS)  
 and  the emitting nebula is accreted towards the  RS.
We adopt the code {\sc suma} which allows  to choose the direction of the clouds relative to the RS.
The  modelling  results which reproduce a  single galaxy spectrum with the highest precision allow to 
classify this  object as an AGN ejecting, an AGN accreting, an SB ejecting or an SB accreting type. 
When more  models are equally valid we suggest that the galaxy is the product of merging.
Our results show that  among the eleven sample  galaxies five  are  such. 
We focus on the N/O  trends with the  oxygen metallicity and with the redshift to identify
the nitrogen/oxygen relative formation processes and the process-rates, respectively, for  intermediate-mass stars.
Our results  show that O/H relative abundances calculated for the sample galaxies are lower than 
 solar by a  factor  $\leq$5.  Yet,  a few values were found above solar.
He/H  were calculated lower than solar by factors $\leq$ 100 and N/H by factors $\leq$135.


}

\maketitle

\section{Introduction}

The exploration of pristine galaxies is one of the  issues concerning  galaxy spectral analysis   
in search  of the Universe conditions at cosmic dawn. 
The first hints were obtained by the interpretation of high  redshift (z$>6$) galaxy spectra 
 when new powerful telescopes e.g. the James Webb Space Telescope (JWST) were installed  
(Vanzella et al 2023, Shaerer et al 2002, 2003, Erb et al 2002, etc ).
Low metallicity is  a basic prerogative of  early time galaxies.
However, the data  concerning  the line fluxes  for each of the observed  objects are  few and hardly
 constrain  the results of  modelling.  
Yet, those obtained by  modelling the line spectra of   low redshift  galaxies by  
applying the direct strong  line method 
to local young, low-mass  and metal poor objects show very low element abundances.
Therefore the  methods adopted for the spectral analysis of  high z galaxies  are  similar to those 
adopted  for  local objects, e.g the strong line  
and the \Te one which are used to calculate  metallicity in terms of the O/H relative abundance from the 
observed data (e.g. Berg et al 2012, Umeda et al 2022, Izotov et al 2021,  etc) in search 
for pristine galaxy residuals.
Nakajima et al (2022) claim that the direct \Te method is not always available due to the faint 
nature of the [OIII]4363 line. 
They investigate optical line metallicity indicators together with fundamental UV properties of 
extreme metal poor galaxies (EMPG).  They  explore the EMPG, in particular 
those with metallicity 10\% solar (12+log(O/H)$<$ 7.69 based on Asplund et al (2009) solar composition).

In this paper we  focus on the  interpretation  of the  galaxy  spectra in  the  local Universe sample  
presented by Nakajima et al (2022) in the 0.00574$\leq$z$\leq$0.05368 redshift range.
This range  is not more special than the others. Yet, this sample adds more data to Contini (2022, hereafter  Paper I)  collection of spectra with [OIII]5007+/\Hb~
line ratios from 3.2 to 10  and [OII]/\Hb~ from 0.2 to 0.85 observed  at low z, which  are generally defined as star-forming galaxies.
However, such line ratios appear also in some AGN spectra.
We will  use for  investigating this sample spectra  ejection and accretion models.  This  approach  to the interpretation of the spectra  offers a  
stronger constraint to model selection.
In particular, we  aim to compare our results with the O/H metallicities calculated by  
Nakajima et al who   deduced that  12+log(O/H)  is within  6.9 and 8.9. 
We  would like also to investigate the trends of the  physical and chemical parameters throughout the redshift  
range reported  by Nakajima et al. 
To  analyse this issue  the code {\sc suma} which accounts  for the coupled effect of photoionization 
and shock (Contini \& Aldrovandi   1983, Viegas \& Contini 1994 and references therein)  
is adopted.
The results obtained for local galaxies  by {\sc suma} for a number of samples selected  by Contini (2022)
 show O/H  abundance ratios from near-solar   to lower than solar  by factors $\leq$2. 
Much lower O/H 
were calculated for the EMPG sample by the strong line method by Nakajima et al  reaching 10\%  of the 
solar value.
The He/H relative abundances  remained  uncertain (e.g. Paper I, Dittman et al 2001, Hunger et al 1999,  
Do et al 2018, Isotov et al 2021).  
It  is still an open question which needs   special care therefore
we will  investigate the role of the weak lines e.g.  HeII 4686 and [OIII]4363 in constraining the models.
 We  suggest that by the  detailed modelling of all the line ratios presented by the observations within single spectra the
 full picture of the physical characteristics and of the element  abundances throughout the  galaxy 
 gaseus clouds can be reached.
The distribution of the line ratios -  observed from pristine galaxies - throughout the classification schemes 
(Kauffmann et al 2003, Kewley et al 2001) adopted  for the local ones with similar spectral characteristics 
was  also discussed by Nakajima et al (2022).
This modelling  method which is based on  galaxy samples rich in number of objects
 gives   results on a large scale  (e.g. Berg et al 2016). 

We adopt the code {\sc suma} because it  allows  to choose the direction of the clouds relative to the RS.
The results  are obtained in particular for the two cases,
the emitting nebula is  accreted towards the  RS  and  the emitting nebula
is ejected outwards from the galaxy RS, i.e. 
both the photoionizing radiation flux and the shock  act on the same edge of the emitting clouds or on the opposite 
edges, respectively, leading to different ion distribution profiles between them.
The cloud motions  which  follow   from the interaction  with the interstellar  medium (ISM) 
of star formation events relative to the RS location in the galaxy discs or in the presence of winds and 
jets (Sancisi et al 2008, 
 Murray \& Chang 1997, etc) result in accretion or ejection modes.
Zhang et al (2023)  claim that the gas is expelled from the  galaxies by winds and jets  and 
by galaxy merging processes at 
earlier z, while Ginolfi et al (2023) on the basis of  the ALMA data discussed  gas outflow for metal 
enriched gas   and for  e.g. the AGN A1689-zD1 at z$>$7.

Accretion onto  galaxy disc  such as gas inflow from the circum galactic medium and merger  events 
could   generate large scale shocks (Goldman 2024) and turbulence in the disc outskirts (Goldman \& Fleck 2023).
With regards to AGN, Dittmann, Cantiello \& Jermin  (2021) found that
tidal effects from the supermassive black hole (SMBH) significantly alter the evolution 
of stars in AGN discs.
They claim that "in addition to depending on the ambient 
density and sound speed, the fate of stars in AGN discs depends sensitively on the distance 
to and mass of the SMBH. Not only the location in the disc in which 
stellar explosions occur is affected
but different types of chemical enrichment take place."
Moreover, there is a SMBH maximum luminosity above which the outward pressure exceeds the 
inward pull of gravity.
Concluding, it is suggested that accretion and ejection are  related to the element  abundances.
We will check by modelling the single spectra whether accretion or ejection  are more appropriated 
to each galaxy of the Nakajima et al sample.  
The spectra presented by Nakajima et al (2022)  for  a sample of galaxies at z between  0.00574 and 0.05045
are modelled in Sect. 2.  We will proceed by  classifying 
the  sample galaxies on the basis of the models which best reproduce the observation data.
The results are discussed  considering the physical and chemical parameter trends as function of z  
in Sect. 3.  Concluding remarks follow in sect. 4.

\section{Modelling  the Nakajima et al (2022) sample}

\begin{table}
\centering
\caption{Selected galaxies from the Nakajima et al. (2022)}
\begin{tabular}{lccccccccccccc} \hline  \hline
\ galaxy          & ID & z       \\ \hline
\ HSC  J0845+0131 & N1 & 0.01333   \\
\ HSC  J0935-0115 & N3 & 0.01621 \\
\ HSC  J1237-0016 & N5 & 0.05045 \\ 
\ HSC  J1401-0040 & N6 & 0.01168 \\
\ HSC  J1407-0047 & N7 & 0.05368   \\
\ HSC  J1411-0032 & N8 &0.02617  \\
\ HSC  J1452+0241 & N9 & 0.00574 \\
\ SDSS J1044+0353 & N10 & 0.01317\\
\ SDSS J1253-0312 & N11 &0.02301 \\
\ SDSS J1323-0312 & N12 &0.02275 \\
\ SDSS J1418+2102 & N13 &0.00889  \\ \hline
\end{tabular}
\end{table}

Kojima et al (2020) presented data for 10 EMPG, three of which were selected from the 
Subaru/Hyer Suprime-Cam (HSC) Subaru strategic Program (SSP) catalog and 7 were from the 
Sloan Digital Sky Survey (SDSS). 
Using [OIII] 4363 as electron temperature probe, the EMPRESS team confirmed 2 new EMPG including HSC 
J1631+4426  which shows the lowest metallicity ever reported.
Nakajima et al (2022) paper is part of the EMPRESS program by Kojima et al (2020).
Another spectroscopic follow-up by Isobe et al (2021)
for 13 EMPG candidates all of which were HSC-selected, was performed with Keck/LRIS.
In their paper Nakajima et al report the properties of  nine new HSC galaxies  and  four SDSS
(MagE sources). They obtained 12+log(O/H)  between 6.9 and 8.9 corresponding to 0.02--2 Z$_{\odot}$.

The spectra presented by Nakajima et al  show a number of significant lines rich enough to 
constrain the models. 
Degeneracy decreases when  more lines are reported in a spectrum.  There  is a set of selected lines from the
most significant  ones (e.g. [OIII] doublet in the optical range, [OII] doublets in the optical-near-UV and optical-near-infrared ,
HeII, [NII], \Ha, \Hb~ and \Hg) which constrain the model in the optical range. In the UV  the set is composed of UV lines.
This argument is also connected with the choice of the grids of models (see sect. 2.1) which should be not-uniform, in the sense
that in  some critical ranges  some grid- meshes  are reduced  as much as to obtain a better fit  to all the observed line ratios.

In Table 1  we report the galaxies selected from Nakajima et al sample for our investigation.  
HSC J0912-0104 and HSC J1210-0103  were neglected because  the data are not enough to constrain  the models. 
An identification number (N1-N13) is  assigned to each galaxy in Table 1 omitting N2 and N4.
The  results  reported by Nakajima et al are calculated by radiation dominated models  which  reproduce 
the observed  spectra by satisfactory precision focusing on the strong lines.
The  origin of a line strength can be the same for the oxygen ones as well as  for the other elements.
Quoting Kewley \& Dopita (2002) "oxygen is used as a reference element because it is relatively abundant, it emits strong lines in the
optical regime, it is observed from several ionization-level lines" and  because it has a high collision strength
(Osterbrock 1974).  The weakest lines generally have low collision strengths (e.g. [OIII]4363), they depend  strongly  enough on the
temperature of the emitting gas, and/or (e.g. [OII]3727+3729) they have low critical densities for collisional deexcitation.

The \Te method  is adopted  by the majority of the authors  who calculate in particular the  
O/H  metallicity. 
We  intend to  reproduce  as close as possible all the lines (from different ions of different elements) 
presented by the observations in each spectrum consistently, i.e. with the same model.
The shocks which are invoked to  resolve the hydrodynamical picture, are included in the calculations. 
The set of the input parameters which leads to the most acceptable fit of the whole spectrum represents the 
final model.  We consider that the contribution
of  gas in  different conditions throughout the clouds  downstream  of the shock front should 
be accounted for.  
Here,  compression can be strong and temperatures can be high depending on  the shock velocities. 
The physical conditions can  vary within a large range throughout the clouds when shocks are at work. 
This leads to discrepancies  between the results obtained by the  \Te methods compared to  those obtained by 
our models (see e.g. Paper I).
However our modelling method which selects each model through a large grid requires long time spending 
for each of 
them and it is therefore less  adapted to the huge number of galaxy spectra  provided by  modern surveys.
The sample presented by Nakajima et al (2022) in their Table 1 is   suitable to our investigation.
Nakajima et al neglected  AGNs in their models. We consider  also AGN dominated models in the present 
analysis  and we compare the results with those obtained for SB dominated galaxies.

\subsection{Model calculations: the input parameters }

The code {\sc suma} accounts for photoionization and shock. The input parameters related with the shock 
are : the shock velocity \Vs, the preshock density \n0 and the preshock magnetic field \B0.
   A first hint to \Vs comes from  observed FWHM of the lines, e.g. [OIII]5007,4959 and for \n0 it comes from the observations of the characteristic
line ratios such as [SII] 6716/6730 and/or [OII]3727/3729. However, the [OII] lines are often blended and  [SII] may contain a large contribution
from diffuse  extragalactic gas.
 Then, to obtain   more information about the input parameters which define the  model, we consult the grids
calculated by Contini \& Viegas (2001a,b). From these grids we  can also distinguish
between shock dominated models or radiation+shock coupled models.
We select the initial  parameter set among those  which  yield results as close as possible to the observation data.
On this basis  we start to build a grid. However, the grid- meshes for each of the physical and chemical parameters are  often
modified comparing the calculation  results with the data  until all or at least most of  the line ratios in a single spectrum
are successfully reproduced.  The calculations  combine photoionization and heating by the primary and secondary radiation fluxes 
with collisional ionization and heating.

For  all models \B0=10$^{-4}$Gauss is adopted in the present work.
\Vs and \n0  are  used in the calculations  of the Rankine-Hugoniot equations
  at the shock front and downstream.
They  are combined in the compression  equation (Cox 1972)  which is resolved
throughout each slab of the gas in order to obtain the density profile downstream. 
This is a prerogative of models accounting for the shock.
  The higher \B0, the lower  the compression  n/\n0 $\propto$ (1/\B0).
\B0 is chosen such as to obtain the  density n in the range  observed  in the  SB and AGN emitting clouds.
As  \n0 and \B0  are correlated, the  series of  preshock-densites  which  better reproduce  the spectra of the Nakajima sample
galaxies corresponds to a sequence of frozen-in magnetic fields B= n \B0/\n0.

Primary radiation  from  the starburst  is approximated by a black-body.
  The input parameters  are the effective temperature  \Ts and the ionization parameter $U$. A pure black-body radiation
 referring to \Ts is a poor approximation for a starburst, even adopting a dominant
 spectral type. 
 We follow the Rigby \& Rieke (2004)  advise to adopt black body radiation for SB on the  basis of a large range of ionization
parameters, effective temperatures and also relative abundances of the elements.
Besides the  black-body spectral energy distribution (SED), we use  also power-law SEDs when the radiation flux comes 
from an AGN. Moreover, we combine both 
the black-body and the power-law with collisional ionization  by the shock.
This paper, as well as other papers  which show the results of models of shocks and photoionization, is aimed to  provide an answer
about the role of the shocks in the interpretation of the spectra. Comparison of  the ionization rates for
the primary radiation, for the secondary diffuse radiation and for  collisional  ionization rates for some interesting ions as function
of the temperature in the gaseous clouds are shown by   Contini (1997,  Figs. 3-5).  At T$\sim$10$^5$ K, i.e. in the UV domain,
collisional  ionization rates  dominate.
\vspace{5mm}

The input parameter  which  represents the radiation field  in AGNs is the power-law
flux  from the active centre (AC) $F$  in number of photons cm$^{-2}$ s$^{-1}$ eV$^{-1}$ at the Lyman limit.
The spectral indices are $\alpha_{UV}$=-1.5 and $\alpha_X$=-0.7.

When the primary radiation flux - which depends on the RS - reaches the clouds, it  affects the surrounding gas.
 This  region  is not considered
as a unique cloud, but as a  sequence of slabs with different thickness calculated automatically
following the temperature gradient. The secondary diffuse radiation is emitted
from the slabs of gas heated  by the radiation flux reaching the gas and by the shock.
Primary and secondary radiations are calculated by radiation transfer through the slabs. 
The  line fluxes which appear  within the spectra are calculated  integrating throughout the clouds  on
a maximum of 300 slabs.  

The geometrical thickness of the emitting clouds $D$  and the relative abundances of the elements to H 
are  input parameters.
In our model the line and continuum emitting  regions throughout the galaxy cover  an ensemble of fragmented clouds.
The geometrical thickness of the clouds is   calculated
consistently with the physical conditions and element abundances of the emitting gas.
$D$ is  hardly  deduced from the observations. It  determines whether the models are matter-bounded or 
radiation-bounded.
The shocks can create turbulence directly via the  Richtmyer-Meshkov instability (Meshkov 1969). In addition the shocks can   generate shear flows
which in turn can  create turbulence by shear instability and  by the Kelvin-Helmholtz instability.
The turbulence is expected to be supersonic and could cause fragmentation of matter that leads to 
clouds with 
very different geometrical  thickness  coexisting  in the same  galaxy (see e.g. Fonseca-Faria et al 2023).

The abundances of the elements which enter directly in the calculation of the line intensities, affect 
in different ways the cooling
rate of the gas  downstream   and  consequently they  shape the distribution of the ion fractional abundances  throughout the clouds.
They  are calculated  resolving in   all the slabs  the ionization 
equations for each element (H, He, C, N, O, Ne, Mg, Si, S, Ar, Cl, Fe) in each ionization level.
Finally, the calculated line ratios, integrated throughout the cloud thickness, are compared with the
observed ones. The calculation process is repeated
changing  the input parameters until the observed data are reproduced by the model results,  at maximum
within 10-20 percent
for the strongest lines and within 50 percent for the weakest ones.
 Those thresholds are indicated by the approximation of the  coefficients (for ionization, recombination, dielectric recombination,etc)  
used in  model calculations and  by the observation errors which  can reach 50\% in the  weak lines flux. 
  The accuracy of the chemical abundance determination can be reached depending on  the number of meshes in the grids
and their dimensions. The accuracy of the  calculated coefficients which compose the code   depends on  quantum mechanics.
  A better fit of the spectra can be found by running more models
and building grids with  relatively small meshes.

 The dust-to-gas (d/g) ratio is also an input parameter. It regards in particular the
SED affecting  dust reprocessed radiation and its peak. Moreover, a high d/g can reduce the  
non radiative to a radiative shock through the mutual collisional heating and cooling of dust grains and gas atoms.

\begin{figure*}
\centering
\includegraphics[width=6.6cm]{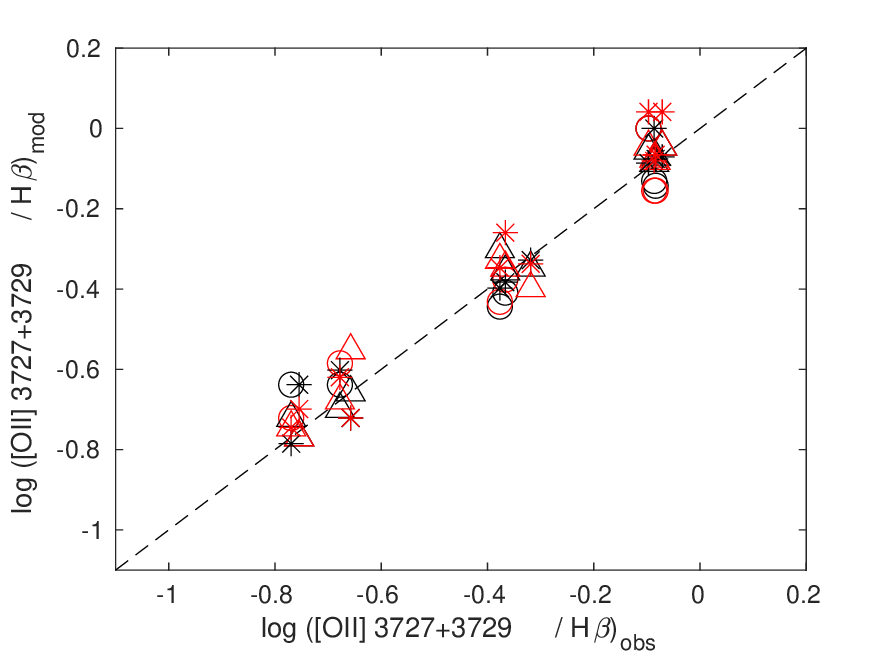}
\includegraphics[width=6.6cm]{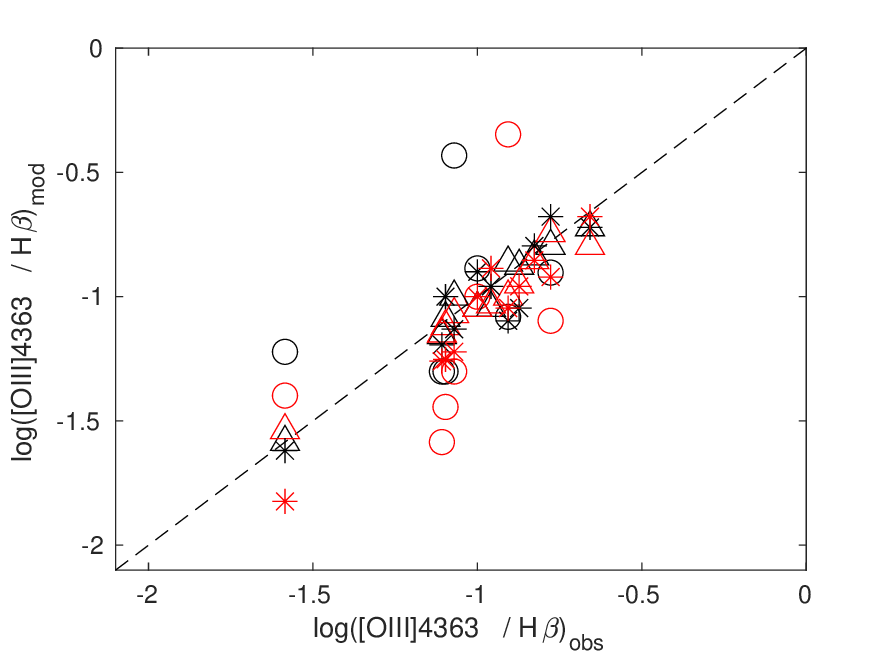}
\includegraphics[width=6.6cm]{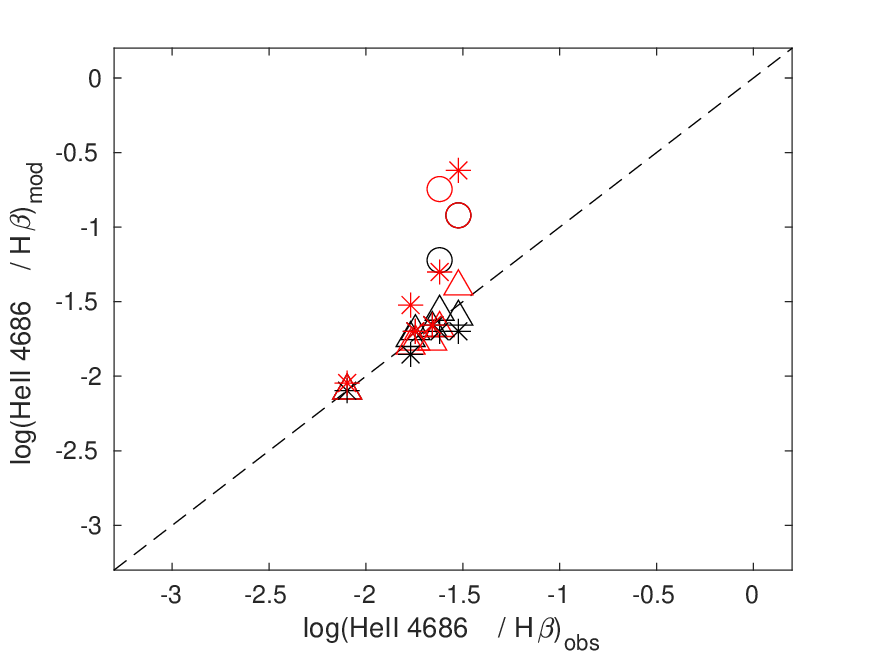}
\includegraphics[width=6.6cm]{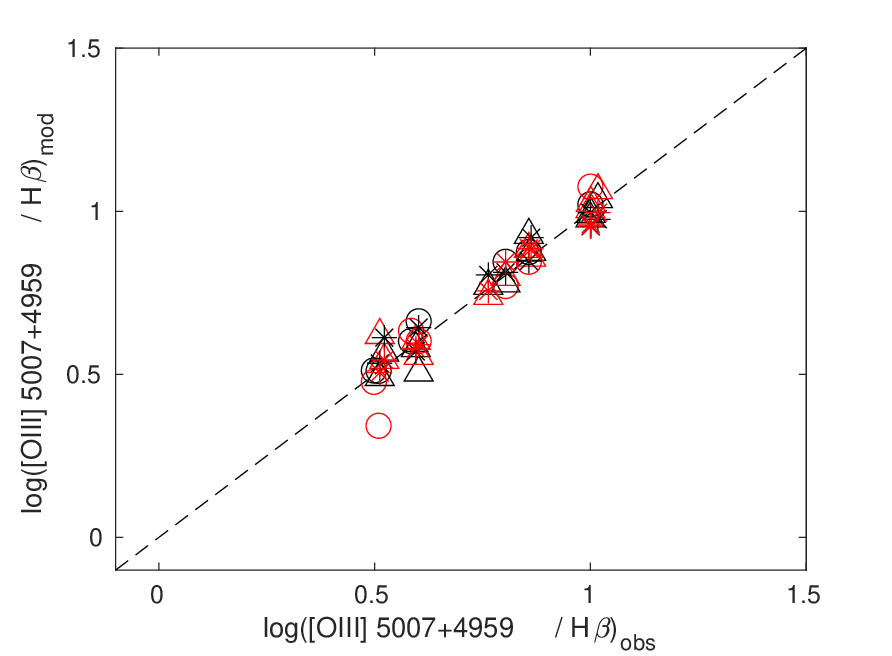}
\includegraphics[width=6.6cm]{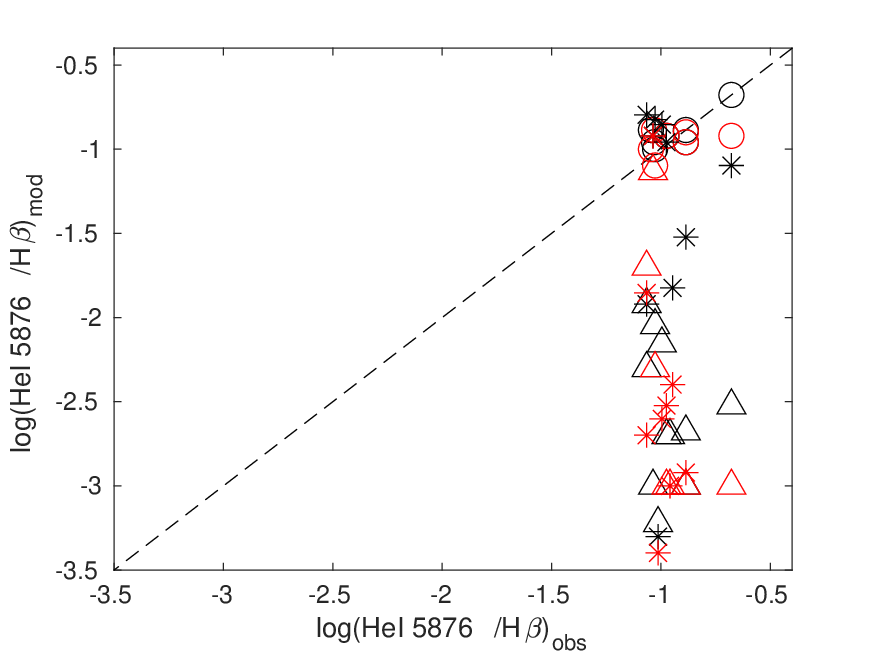}
\includegraphics[width=6.6cm]{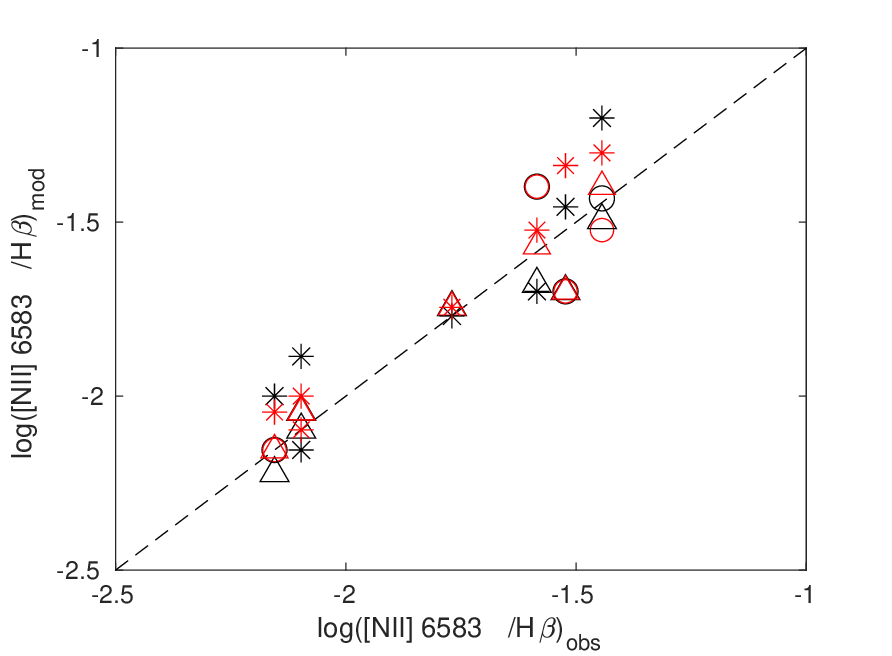}
\caption{Calculated (mod) versus observed (obs) line ratios to \Hb.
Black  symbols refer to accretion, red ones to ejection. Open circles:  models (mp) calculated
by radiation adapted to  an AGN and  solar He/H; open triangles: models (mh) calculated by radiation adapted to AGN and by
different He/H.  Asterix: models (ms) calculated by radiation adapted to starbursts and  different He/H.
 The models are described in the bottom of Tables A.1-A.8}

\label{FigComp}

\end{figure*}

\begin{figure*}
\includegraphics[width=12.0cm]{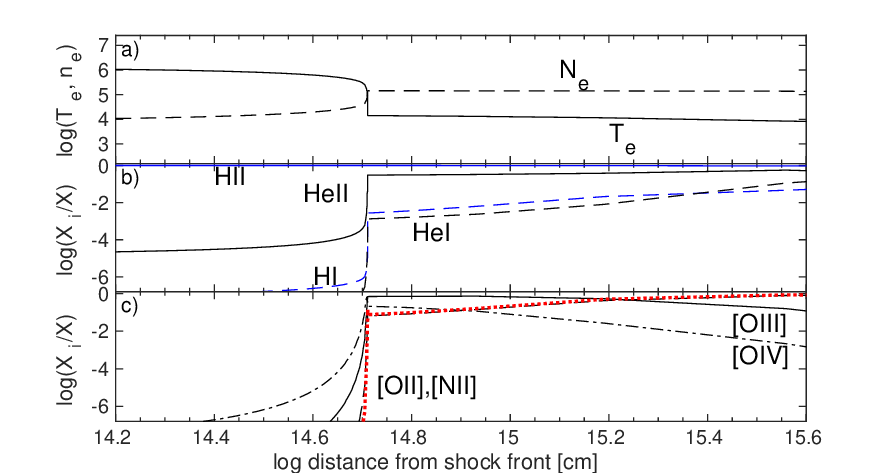}
\caption{Profiles of the physical parameters throughout the N12 galaxy clouds  in the
accretion case: the shock front and the edge reached by the  AC flux are both on the left.
Top diagram:  black lines: \Te (solid),  \Ne (dashed);
middle diagram: blue: HI (dashed), HII (solid); black : HeI (dashed), HeII (solid);
bottom diagram: black lines refer to oxygen ionization stages: [OIV] (dot-dashed),  [OIII] (solid), [OII] 
(dashed); red dotted line: [NII]}

\label{naka120}
%\end{figure*}

%\begin{figure*}
\includegraphics[width=8.6cm]{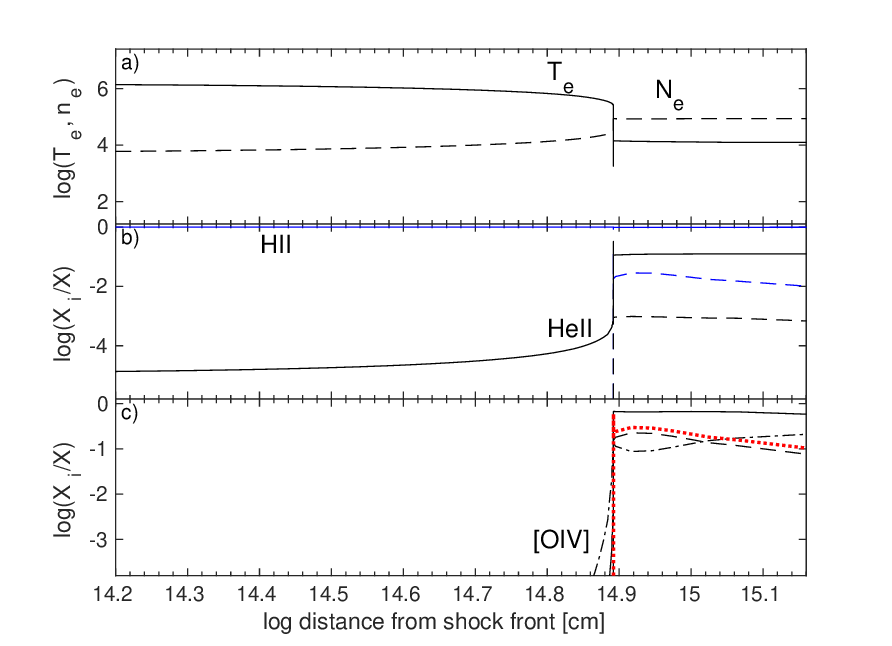}
\includegraphics[width=8.6cm]{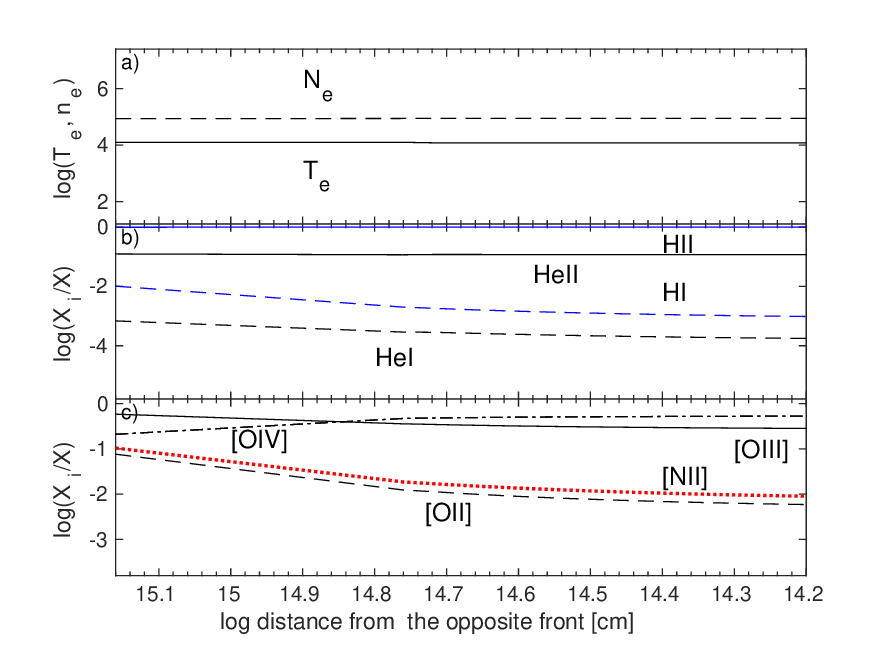}
\caption{The ejection case. The clouds are divided in two contiguous halves which are  displayed by 
the left and right panels.  In the left panel
the shock front is on the left and the X-axis scale is logarithmic. In the right panel
the right edge is reached by the flux from the AC. The X-axis scale is logarithmic
in reverse in order to have the same detailed view of the cloud edges.
Symbols as in Fig.\ref{naka120}.
}

\label{naka121}

\end{figure*}

\subsection{Model calculations: the line ratios}

The  line intensities  calculated in each slab of gas are integrated throughout the clouds taking into consideration 
the radiation flux from the  RS and the shocks.
We first try to reproduce the relatively high  line ratios (e.g. [OIII] 5007+4959/\Hb, [OII] 3727+3729/\Hb) 
 constraining the  physical parameters  
to best fit  all the  available oxygen line ratios including [OIII]4363/\Hb. 
The [NII]/\Hb ~
 ratios are calculated  considering that charge exchange
reactions are at work between O$^+$ and H$^+$ and  between N$^+$ and H$^+$. Therefore,  [NII]/[OII] 
are  adjusted
by selecting the most suitable N/O relative abundance. The  [NeIII]3689+/\Hb~ ratios 
(the plus indicates that the doublet is accounted for) follow the
[OIII]5007+/\Hb ~ ones due to their  atomic structures. They  are regulated by the Ne/H relative abundance. 

The line ratios to \Hb =1 are then compared with the  data and the best fitting  models are selected. They  appear
in Tables A.1-A.8. 
Each galaxy from the Nakajima et al sample is  identified by  the number  which appears in Table 1.
In the following [OIII]5007+ indicates [OIII]5007+4959, [OII] indicates [OII]3727+3729, HeII  indicates HeII 4686, 
[NII] indicates [NII]6583  and HeI indicates HeI 5876.
In row 13 of Tables A.1-A.8 the \Hb~ flux calculated at the nebula is reported, followed in rows
14-18 by the values of the physical parameters adopted for each model and of the relative abundances  
of the elements to H in rows 19-23.
We adopt the simplified picture of  a cloud moving towards the photoionizing source  or in the opposite direction  
as  accretion  or ejection, respectively.  The calculations  translate  to
radiation from  the  RS (such as the AC for AGN) reaching  the very shock front edge of the 
cloud (accretion) or  radiation reaching the edge  opposite  to the shock front (ejection). 
We will  calculate the  line ratios in both cases and we will  try to reproduce the observed ones.
Each  galaxy spectrum presented in  Tables A.1-A.8 therefore  is splitted into   accretion and ejection 
 identified by the parameters str=0 and str=1, respectively, which appear in the bottom row. 

 In particular, in Tables A.1-A.2 the observed line ratios  are compared  with model results mp1.0 - mp9.0 and mp1.1 - mp9.1
(recall that 0 refers to accretion and 1 to ejection). The models account for
the radiation flux from the AC adapted to  an AGN i.e. a power-law and solar He/H.
The other element ratios to \Hb~ are selected among the best fitting ones.
In Tables  A.3-A.5 the observation data are compared with  models  mh1.0-mh13.0 and mh1.1-mh13.1
obtained adopting a power-law flux as for AGN and the  He/H relative abundances  best fitting the observed
HeII4686/\Hb~ line ratios.
In Tables A.6-A.8 the  data are compared with  models  ms1.0-ms13.0 and ms1.1-ms13.1
obtained adopting a  black body radiation flux   as that used for starbursts (see Paper I) and  consistent He/H.

 Fig. 1   confirms in  a graphical way  the results  presented in Tables A.1-A.8.
We neglected the data error bars  both to obtain  a clear picture and  to   easily distinguish the  less fitting
results.
Fig. 1  shows that the  error is minimum for [OIII]5007+ /\Hb, increasing for [OII]/\Hb~ ratios  which are low in the present galaxy sample.
Discrepancies  can  be high  for [OIII]4363/\Hb~ ratios which  are even lower than [OII]/\Hb.

 O/H was found underabundant  by a factor $\leq5$  relative to  solar
((O/H)$_{\odot}$=4.9$\times$10$^{-4}$, Grevesse 2019) in  mh1.1 in N1 and mh3.1 in N3 and higher than solar
in  AGN mh13.1, mh10.1 and in  both SB and AGN   N3, N11 and N8 (see  Table 4.)
Ne/H  relative abundances are lower than solar by 40\%  
((Ne/H)$_{\odot}$=8.5$\times$e10$^{-5}$, Grevesse 2019, Young 2018). 
About  [SII]6716/\Hb~ and [SII]6730/\Hb,
the observed [SII]6717/[SII]6730 (see Tables A.1-A.8) doublet line ratios are generally $\geq$ 1  indicating very 
low electron densities (Osterbrock 1974) which are  less  suitable
to galactic clouds. The disagreement with model results  which show ratios $<$1 is explained by the 
contribution of diffuse intergalactic gas (Paper I).
In fact, the first ionization potential of S is 10.31 eV, lower than that of H (13.51 eV) and the [SII] lines 
can be strong in the ISM where  temperatures  and densities are relatively low (T$\leq$ 10$^4$K, \n0$<$ 10 \cm3).
The hydrogen line ratios \Ha/\Hb~ and \Hg/\Hb~ are calculated in each slab.  
\Hg 4340  can be blended with [OIII]4363 and \Ha~ is blended with the [NII]6548, 6583 doublet.
\Ha/\Hb~ ratios are used to correct the spectra from reddening. The values are  generally fixed to $\leq$3 
considering $T\leq$10$^4$K. However models accounting for the  shocks lead to large regions of gas at 
higher temperatures therefore model results show  also \Ha/\Hb  $\geq 3$.

\subsubsection{Helium lines}

We  focus on the He lines following the results discussed in Paper I.
In Nakajima et al (2022, their table 1) HeII 4686/\Hb~ lines  are  weak compared to those  in AGN spectra 
which  can be strong (Aldrovandi \& Contini 1985).  The HeI 5876 lines  on the other hand
 are  $\sim$ 0.1 in the  spectra from different objects.  
The HeII/\Hb~ ratios in the N1, N5 and N9 spectra are upper limits  and in N7 and N8 are not provided by the data.
We adopted  solar He/H in the first modelling trial of HSC galaxies (N1-N9) because it  generally fits the 
HeII/\Hb~ and HeI/\Hb~ line ratios observed from  various objects at  different redshifts (Tables A.1 and A.2). 
We realized that  solar He/H ((He/H)$_{\odot}$ = 0.085 from Grevesse 2019) was too high  and that 
for nearly all the objects the  HeII/\Hb~ line ratios  calculated by models mp1.0 - mp9.0 and mp1.1 - mp9.1
overpredicted  at least by  a factor of 10 the observed values, while the calculated HeI 5876/\Hb~ line 
ratios nicely reproduced the data.  Therefore, we repeated the  modelling of the spectra of all the galaxies by more 
adapted  He/H relative abundances by models mh1.0-mh13.0, mh1.1-mh13.1, ms1.0-ms13.0 and ms1.1-ms13.1 (Tables A.3 - A.8).

 HeII/\Hb~ line ratios in the  spectra  reported by Nakajima et al
 are reproduced by AGN dominated  models with He/H lower 
than solar  by factors between  $\sim$ 100 for mh12.1 and 1.7  for mh7.1 (Tables A.3-A.5). 
Solar and higher than solar He/H were certainly not found for some galaxies spectra where the 
observed HeII/\Hb~   were low or absent.  
Moreover, low HeII/\Hb~ line ratios cannot result from models which adopt 
 the relatively strong  radiation flux  which yields [OIII]/[OII] ratios ranging between $\sim$5 and $\sim$60.
 Therefore the parameter which can reduce strongly the HeII line intensity is the He abundance.
We had to recalculate all the other spectral line ratios  in order to reproduce the observed  values because He is a 
strong coolant.
However, the satisfactory fit found for  HeI 5876/\Hb~  for nearly all the spectra   by the mp1-mp9 models 
(Tables A1-A2)  was lost.  This can be seen in  Fig. 1 where we compare the observed with the calculated line ratios.
The dilemma between an acceptable HeII 4686/\Hb~ or an acceptable HeI 5876/\Hb~ was resolved by  arguing that
HeI is  a permitted  recombination line therefore the contribution from the ISM 
to the observed  line is sensible.  This hypothesis is strengthened by the fact
 that HeI/\Hb~  ratios vary within a small range (Fig. 1, left bottom diagram) in  galaxies with different spectra 
 because the diffuse ISM  has temperatures T$\leq$ 10$^4$K.
Therefore, in the following we will adopt  reduced or enhanced He/H relative abundances in order to  reproduce the HeII4686/\Hb~ 
line ratios neglecting   the HeI/\Hb~  fit for both AGN and SB models.
The  results of HSC spectra  reported in Tables  A.1-A.2  are constraining enough  to support this discussion and 
the SSDS spectra will be properly calculated. The results are reported in Tables A.3-A.8.

Model results for SB galaxies appear in Tables A.6-A.8. They show that the minimum He/H  results by a factor 
of 85 lower than solar for
for both ms12.1 and ms12.0 models. He/H is solar for model ms6.0 but  this model does
 not reproduce satisfactorily the HeII/\Hb~ line  ratio. A good fit of both HeII/\Hb~ and HeI/\Hb~ is obtained  
 by model ms13.0  with He/H =0.09, but only in the accretion case.
A higher than solar He/H by  factors of $\sim$3 and $\sim$10   reproduces reasonably the HeII/\Hb~ line ratios in the ms11.0  and ms6.1 
spectra, respectively, but HeI 5876/\Hb ~ line ratios result overpredicted by the same factors. 

Energy loss by  line  emission affects the cooling rate of the gas downstream, therefore  changing the element 
abundances  the physical parameter sets  require remodelling.
Low  abundances, in particular for He which is generally a strong coolant
lead to a reduced cooling rate  and affect all  the calculated line ratios in different ways. 
For the Nakajima et al  spectra however we found that He is often so underabundant  that  increasing or decreasing He/H
 by factors  $\leq$ 10 will not  alterate the line ratios from the other  elements. As a consequence  HeII/\Hb~ 
 ratios can be readjusted  simply by  modifying  He/H.

 \subsubsection{Nitrogen lines}

 The same arguments as those adopted for HeII/\Hb~  can be adapted to  [NII]/\Hb~  because
N/H relative abundances were also found lower than solar ((N/H)$_{\odot}$= 6.76$\times$10$^{-5}$ 
Grevesse 2019) by factors $>$10.
A perfect fit of the [NII]6583/\Hb~ line ratios is meaningless because the  [NII] lines are  extracted from the [NII] 
doublet  6583, 6548. They  can be blended 
 at  high velocities and they are generally blended with the \Ha~ line for \Vs~ adapted to the AGN NLR and the SB.
 The line ratios depend on the  electron
 temperature and density  that are not constant throughout the emitting clouds when shocks are at work. 
 Therefore the error can be sound.
[NII]/\Hb~ line ratios in all the  present observed spectra are low relative to   those shown in other galaxy samples 
at different z. If this depends from a low nitrogen  abundance, as we have explained for He, N will  end its role 
as  a strong coolant in the downstream region.  Consequently, N/H will directly result  by reproducing the observed  
[NII]/\Hb~ line ratios. 

\subsection{Distribution of \Te, \Ne and of  oxygen ion fractional abundances downstream}

The line  ratios within each spectrum can be understood following the profiles of the physical parameters
throughout the clouds. We present for instance in  Fig. 2  and Fig. 3  the profiles 
of \Te, \Ne and of the
main ion fractional abundances in the N12  galaxy clouds for accretion and ejection models, respectively, relative to an  AGN.
In models which account for the shock,
immediately behind the shock front, the gas is thermalized to
a temperature  T=1.5$\times$10$^5$(\Vs/[100 \kms])$^2$ K.
At  high temperatures e.g. for \Vs=300 \kms~ T$\geq$10$^6$ K, recombination coefficients are very low and the 
cooling rate of the gas is low.  At T between 10$^4$ K and 10$^5$ K  UV lines and  coronal lines
in the IR are strong and lead to rapid cooling and compression
of the gas. If the cooling rate is so high as to
drastically reduce the  temperature  eluding intermediate ionization-level lines,
the calculated spectra will be less adapted to  reproduce the observed ones because  intermediate 
ionization-level lines (e.g. [OIII], [OII] etc) will be very weak.
However, the  temperature after the  drop is maintained at 10$^4$K by  secondary diffuse radiation which comes from
the slabs of gas heated by the shock and by the photoionizing flux. At T $\sim$ 10$^4$K the gas (in particular oxygen)
shows strong lines from the intermediate-ionization levels.
It can be noticed that the models are matter-bounded, as it is the case for  most of the galaxies where $D$ is
 small ($<$0.01 parsec).  Fig. 2 and Fig. 3  show that HeI lines are very weak.
and some other recombination lines  disappear.
We have added in the figures the fractional abundance of O$^{3+}$/O although [OIV] lines were  not observed by Nakajima et al
to show  the region of hot gas downstream where high-ionization-level lines  up to  e.g. the Fe coronal ones can be strong.
 The two phase regime which appears in Figs. 2 and 3  explains the  observed spectra containing both relatively 
strong high ionization-level lines and the usually
strong intermediate-level optical lines emitted  from  specific locations in a galaxy (e.g. Fonseca-Faria et al 2023).

\section{Results}

In  previous sections we  have found that  spectra as those presented by Nakajima et al, i.e.  showing
[OIII]5007+/[OII]3727+ $>$4.8, [OIII]4363/\Hb~ between 0.08 and 0.16 and HeII/\Hb~ between  0.008 and 0.03, 
can be reproduced by strong photoionizing fluxes with a maximum of 2.2$\times$10$^{12}$ photons cm$^{-2}$
s$^{-2}$ eV$^{-1}$ at the Lyman limit as for the AGN model mh5.1 (Table A.3) and a maximum $U$=0.6 as for the SB model ms12.1 (Table A.8) 
and a maximum   \Ts=180000 K as  for the  SB model ms11.1 (Table A.8). Moreover,  relatively low O/H with  a minimum of  0.0001  
by number (12+log(O/H)=8.0) as for the AGN model mh1.1 (Table A.3) and a minimum of 8.36 as for the SB model ms12.0 (Table A.8) derived  
from the calculations.
In general,  low He/H  with a minimum of 0.0008 by number for the  AGN model mh12.1 (Table A.5) and a minimum of 0.001 by number for the SB
model ms12 (Table A.8) both accretion and ejection resulted  by modelling the spectra. Upper limits  were also used to constrain the models.
For all the sample galaxies 
 shock velocities   similar to those  which characterise the narrow line region (NLR) of Seyfert 2 galaxies (\Vs = 100-360 \kms)
 and also to those (\Vs =160-440 \kms) used to model SB galaxies at low z (Paper I) were revealed.
 The calculated preshock densities  were found relatively high with a peak of 4000 \cm3 for the AGN model mh3.0 (Table A.3) and of 3200 \cm3 
 for SB ms12.1 (Table A.8) in order to keep the [OII]/\Hb~ line ratios low enough.
 The [OII] lines have critical densities for deexcitation (\Ne$_{cr}$$<$ 1000 \cm3) lower than for the [OIII] lines
 (\Ne$_{cr}$ $\sim$ 3$\times$10$^5$ \cm3). However, for N5, N7, N8, N11, N12 AGN ejection models the densities 
 are similar to those  of Seyfert 2 galaxies. In the accretion case we can accept high preshock densities because the 
 shock front  reaches the inner regions 
 of the NLR at a   relatively small distance from the AC  confirming that the density trend decreases outwards  throughout the NLR.  
Moreover high densities downstream within the emitting clouds yield the  sharp temperature drop to 
T $\leq$ 10$^4$K which is followed by  gas
recombination  to  intermediate-ionization levels. They are adapted to the  [OII], [OIII], [NII], [NeIII] etc line emission.

\begin{figure*}
\centering
\includegraphics[width=6.6cm]{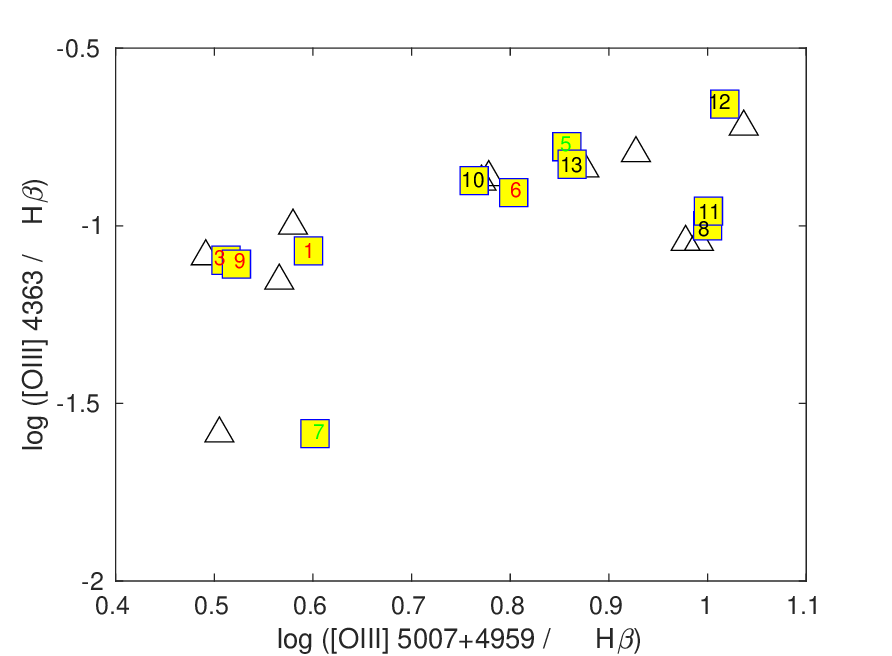}
\includegraphics[width=6.6cm]{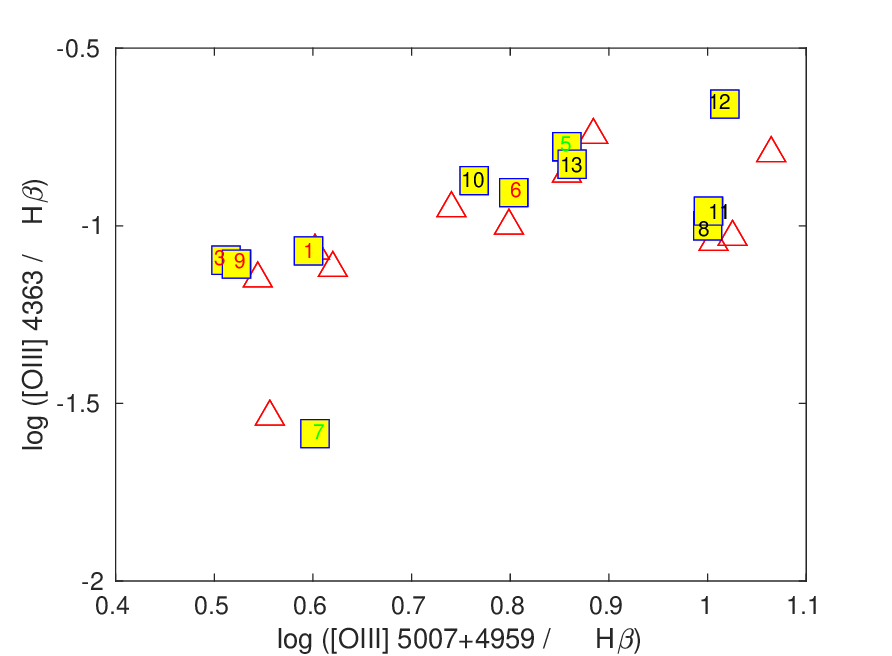}
\includegraphics[width=6.6cm]{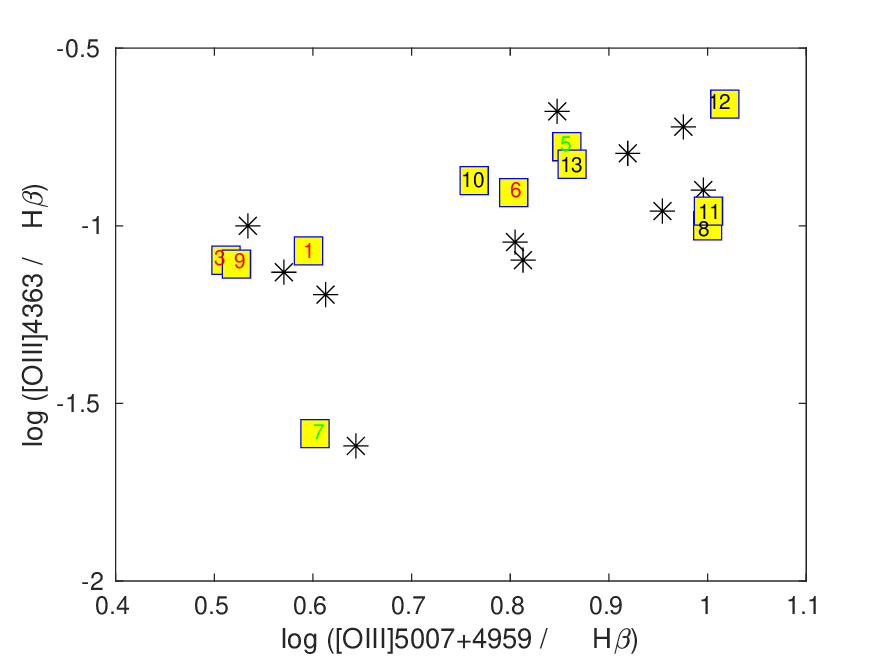}
\includegraphics[width=6.6cm]{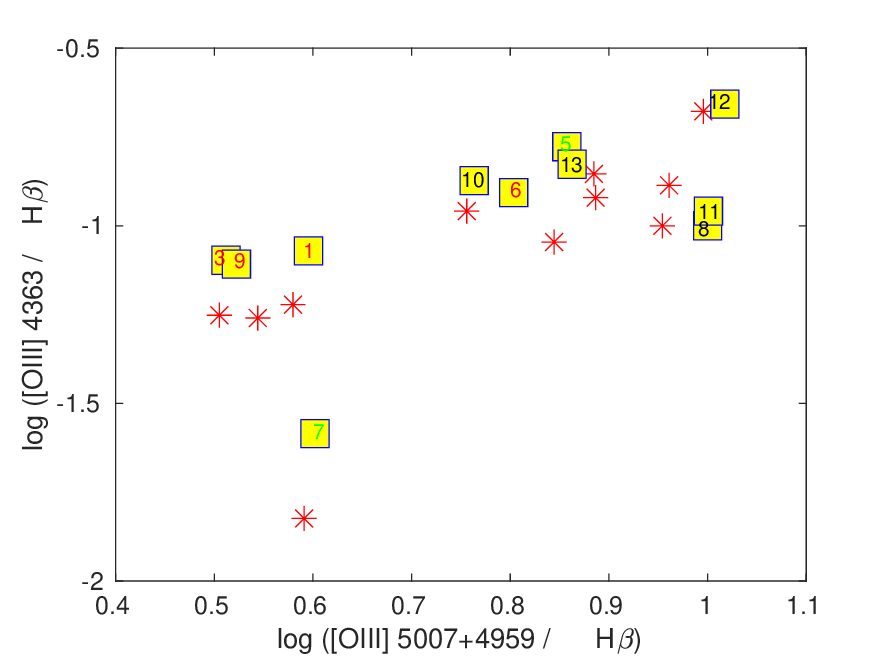}
\caption{
[OIII]4363/\Hb~ versus [OIII]5007+4959/\Hb. 
Yellow filled squares represent the observed values. Other symbols as in Fig.~\ref{FigComp}~  
	Red numbers: AGN; green numbers:  SB; black numbers: merging  galaxies (see sect. 3.2).
}

\label{FigO4O3}

%%%\end{figure*}

%\begin{figure*}
\centering
\includegraphics[width=6.6cm]{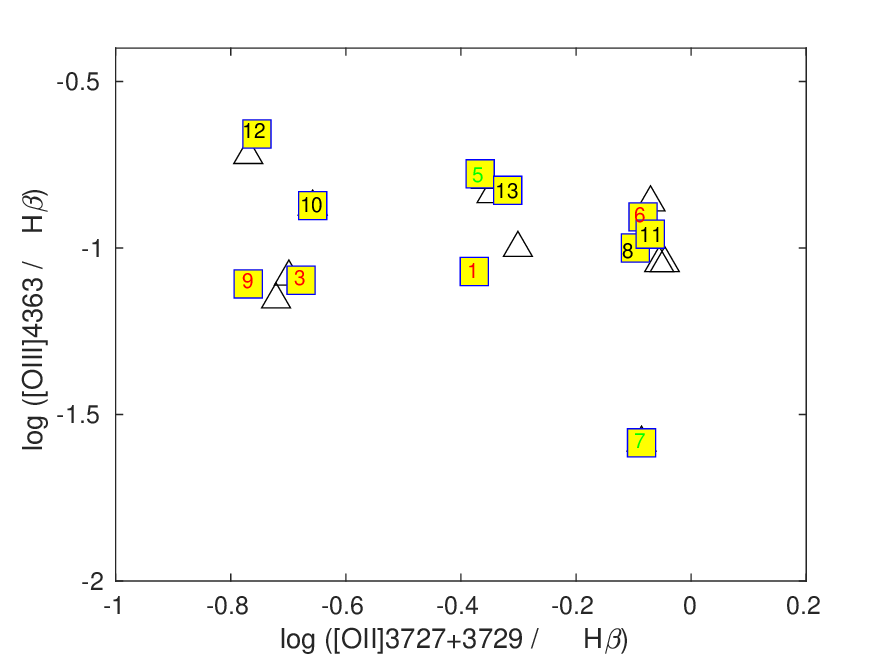}
\includegraphics[width=6.6cm]{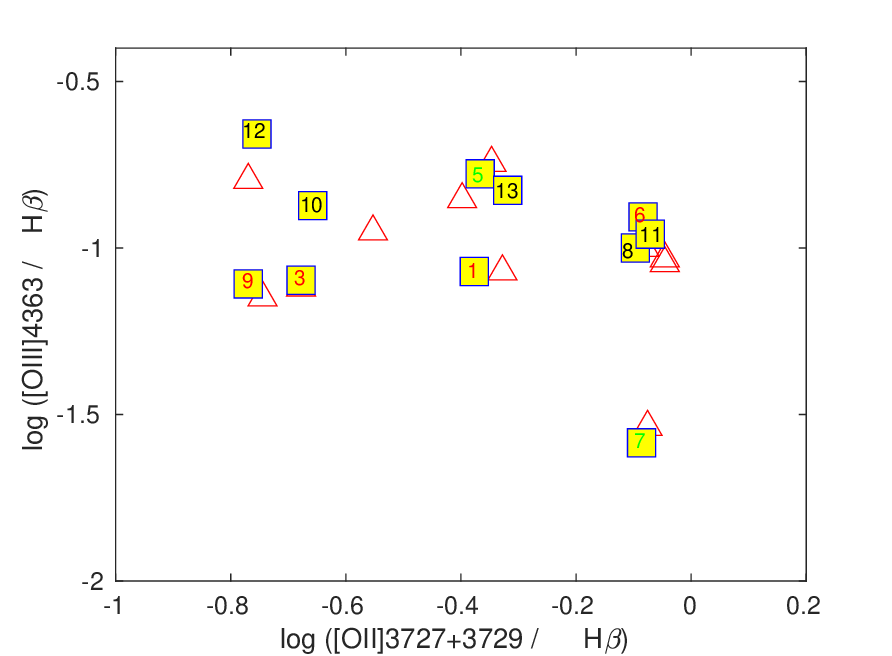}
\includegraphics[width=6.6cm]{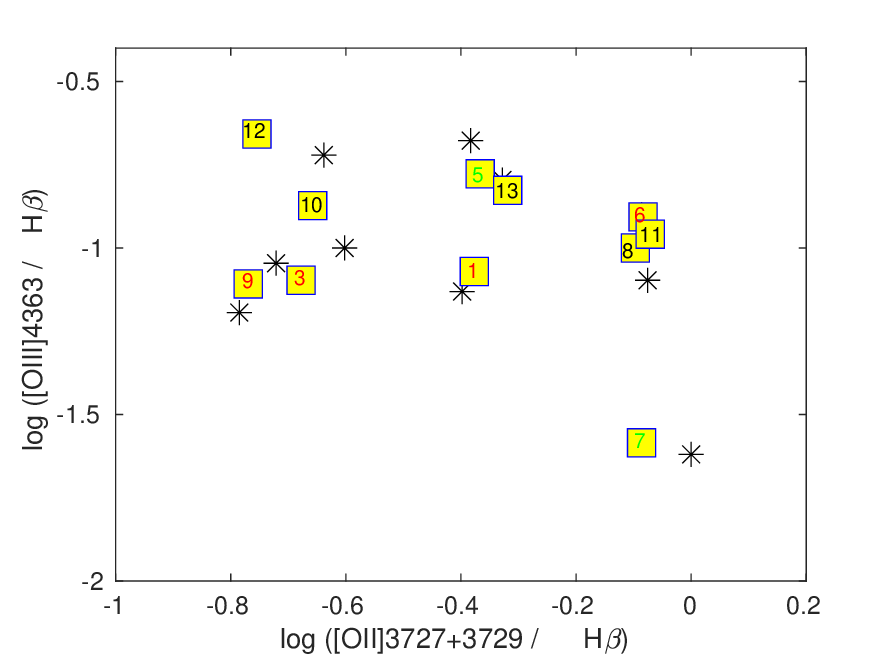}
\includegraphics[width=6.6cm]{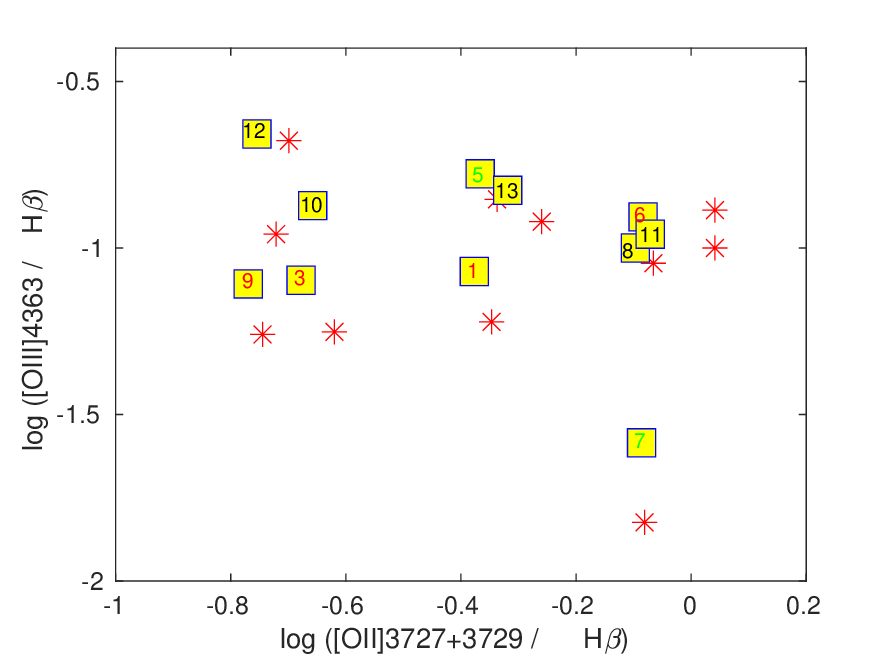}
\caption{[OIII]4363/\Hb~ versus [OII]3727+3729/\Hb.  Symbols as in Fig.~\ref{FigO4O3} }

\label{FigO4O2}

\end{figure*}

\begin{figure*}
\centering
\includegraphics[width=6.6cm]{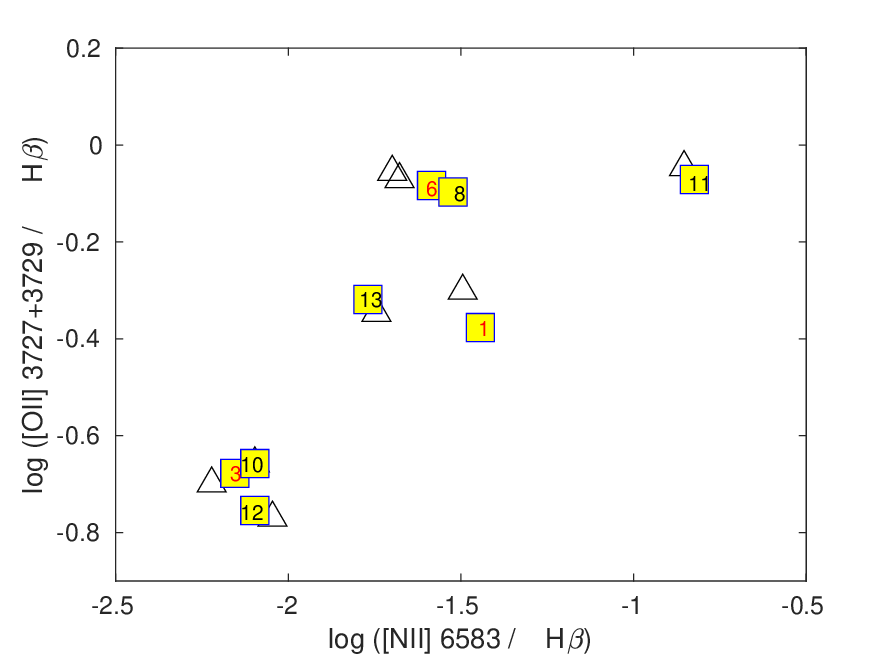}
\includegraphics[width=6.6cm]{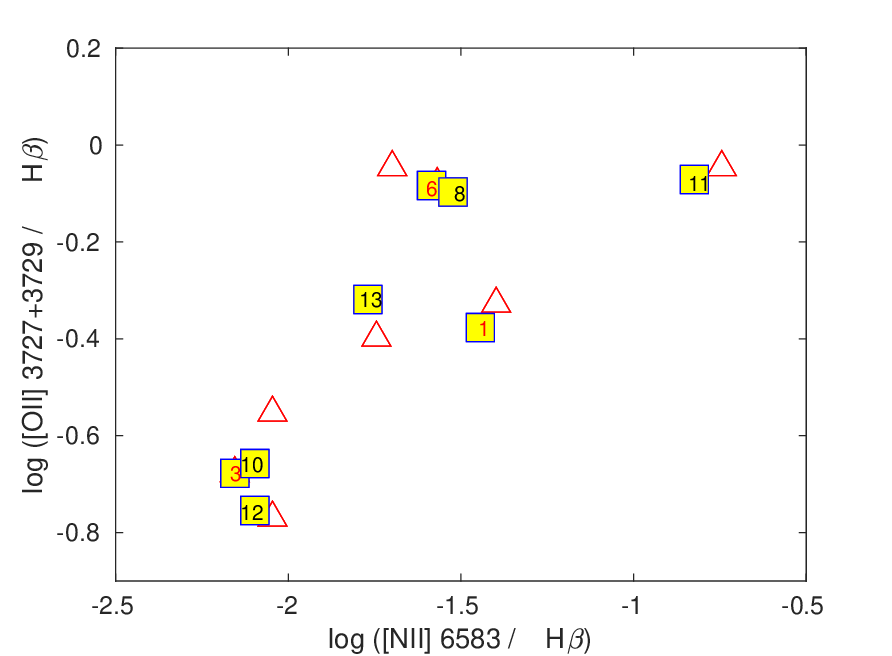}
\includegraphics[width=6.6cm]{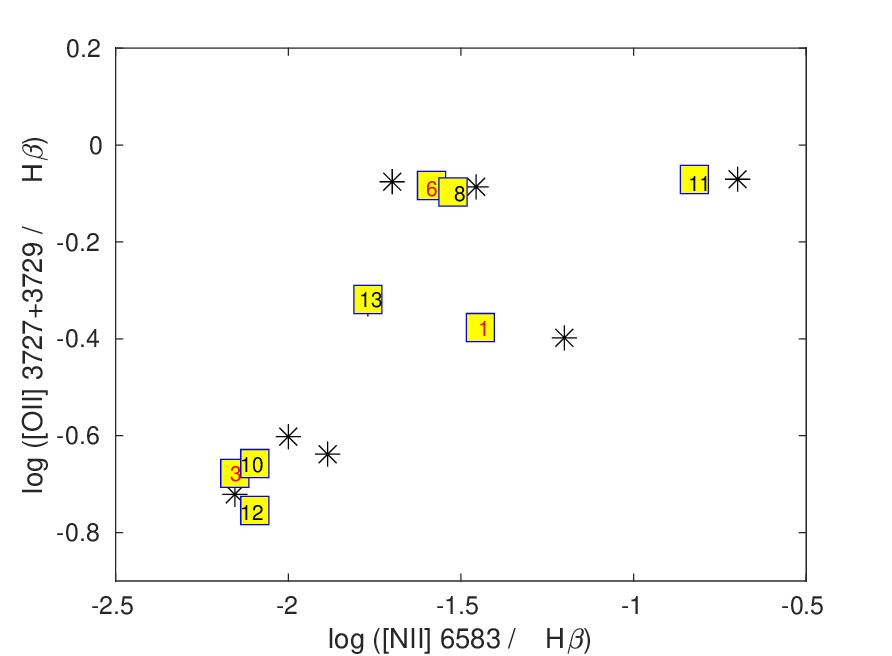}
\includegraphics[width=6.6cm]{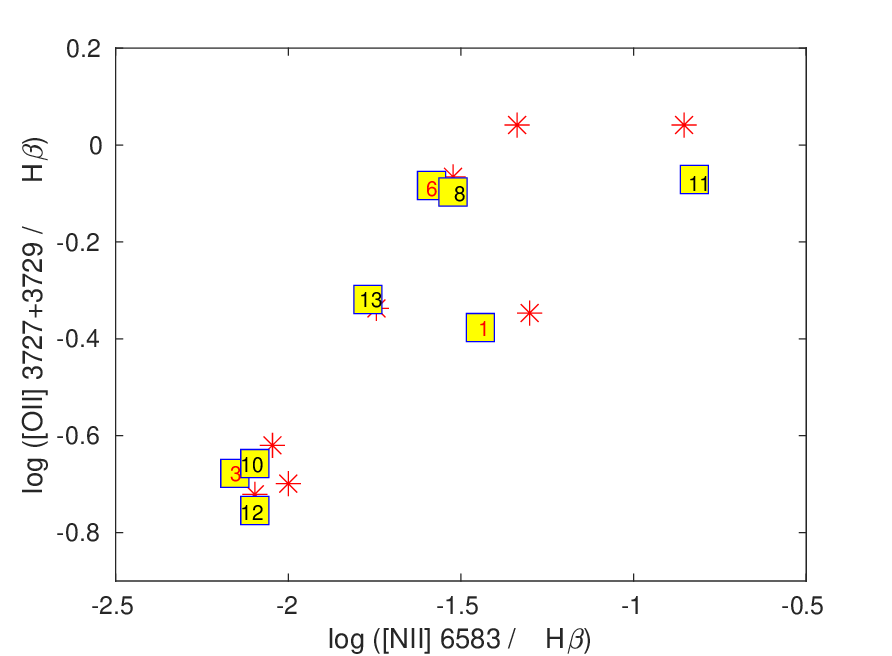}
\caption{[OII]/\Hb~ versus [NII]/\Hb. Symbols as in Fig. 2 }

\label{FigN2}

%\end{figure*}

%\begin{figure*}
\centering
\includegraphics[width=6.6cm]{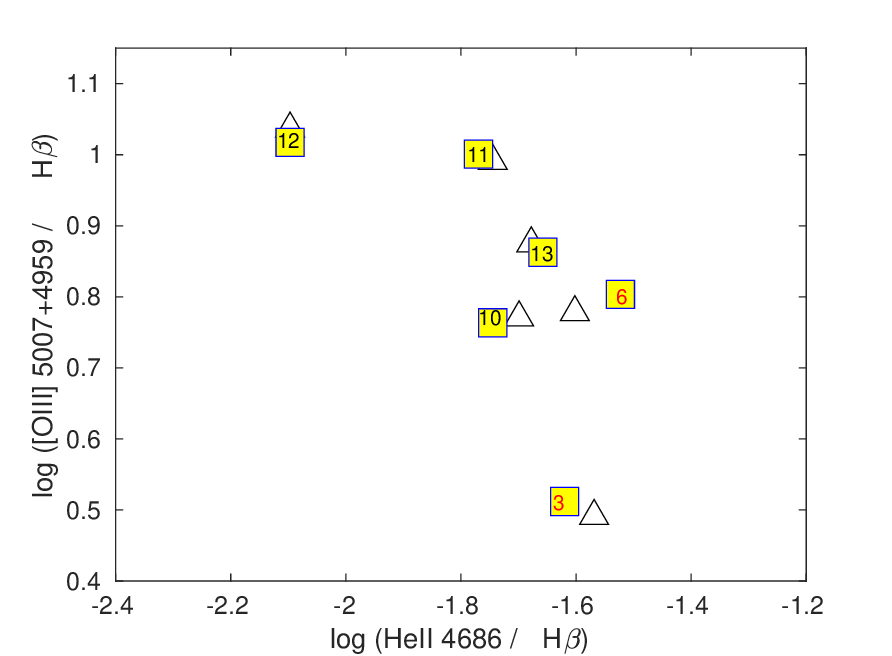}
\includegraphics[width=6.6cm]{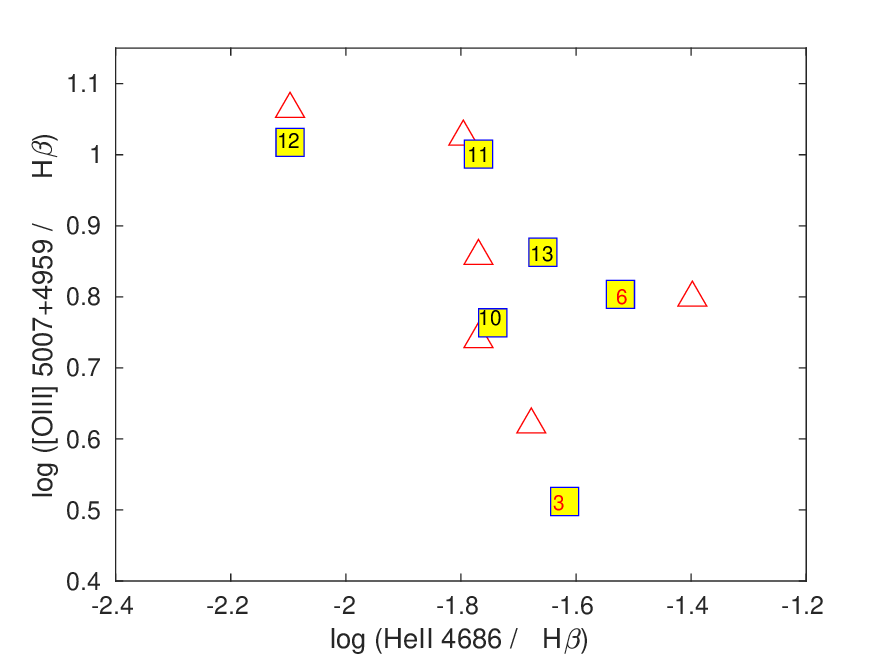}
\includegraphics[width=6.6cm]{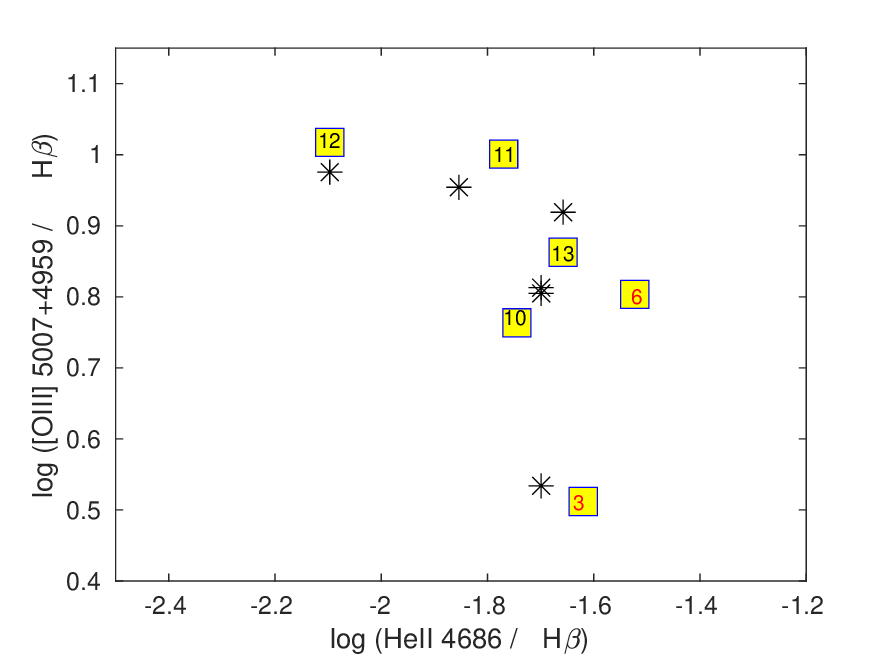}
\includegraphics[width=6.6cm]{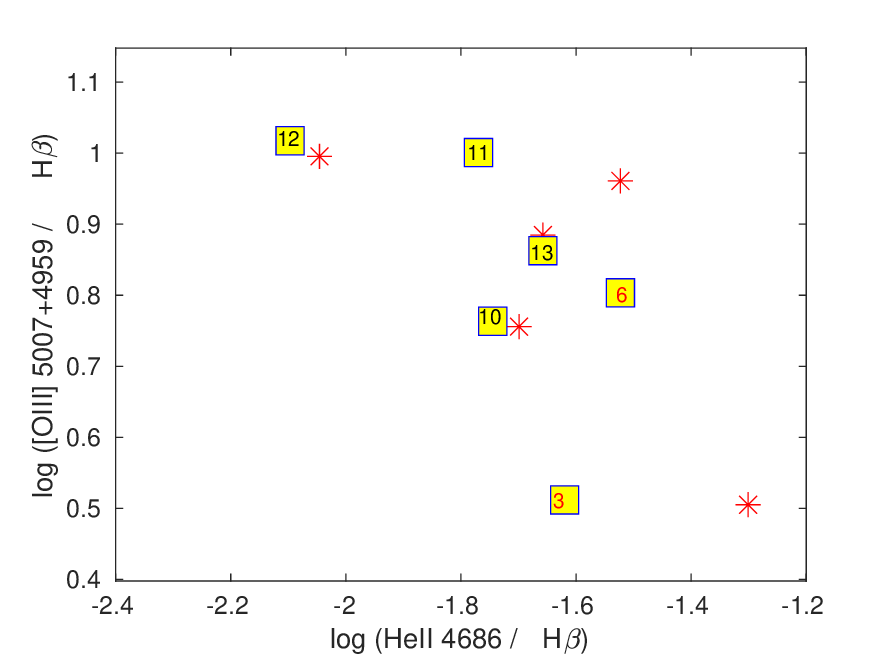}
\caption{[OIII]5007+4959/\Hb versus HeII4686/\Hb. Symbols as in Fig. 2}

\label{O3He2}

\end{figure*}

\subsection{Selecting the galaxy types}

In this work, besides the identification of  each  object  as an AGN or as  a SB, we also distinguish between  
accretion and ejection  by considering the  best approximation (in terms of the least error) 
in  reproducing  the observed line ratios. The most significant lines  are   employed.
Model results are compared with the data which  are reported in Tables A.3-A.8 for  AGN accretion, 
AGN ejection, SB accretion and SB  ejection.
The same  results are illustrated in Figs. 4-7. The data  and model results for AGN models are shown in the top diagrams of 
each figure  and  SB models in the bottom ones.
The accretion models appear in the diagrams on the left  and the ejection models in the diagrams on the right.
We select the best fitting model for  each  galaxy by    neglecting high discrepancies from the data (Tables A.3-A.8). 
We can have a first  hint from  Figs. 4-6 diagrams which, however,  have different axis-scales and can  give
only gross  estimations.  

In Figs. 4 and 5 diagrams the calculated  and observed line ratios are compared. We consider  only those involving oxygen lines to 
\Hb~ (i.e. [OIII]4363/\Hb~ versus (hereafter vs) [OIII]5007+/\Hb~   and 
[OIII]4363/\Hb~  vs [OII]3727+/\Hb) for single galaxies, respectively.
The [OIII]5007+/\Hb~ vs [OII]/\Hb~ diagrams are omitted because all  the models  
reproduce  the line ratios within 10\% for each  spectrum and therefore  they are not constraining. 
In Figs. 6 and 7 diagrams we consider line ratios to \Hb~ from different elements,
[NII]/\Hb~ vs [OII]/\Hb~  and  [OIII]5007+/\Hb~ vs HeII/\Hb~ , respectively.
[NII] and [OII] lines are linked by charge exchange reactions (see Figs 2 and 3) therefore  N/H and O/H 
relative abundances play an important role in  determining  the relative line intensities. 
The scattering of the  galaxies in Fig. 6, however, shows that the observation and  the  model trends are dictated 
also by the physical parameters.
Another interesting  trend is that of  [OIII]5007+/\Hb~   vs HeII/\Hb~  in Fig. 7 diagrams which  is well defined
even  by a few objects i.e. eliminating from the figure HeII/\Hb~ upper limits (for N1, N5 and N9) and the galaxies
where the line ratios  were not provided by the observations (N7 and N8).
Fig. 7 diagrams show that [OIII]5007+/\Hb~ decreases as HeII/\Hb~ increases
In AGN galaxies the [OIII]5007+4959 doublet is
generally the strongest  one in the optical range (see e.g. Dors et al  2021). The [OIII] doublet  as well as  HeII   which is a 
relatively strong permitted  line  
depend on the  radiation flux from the AC and on the temperature of the emitting gas.  This is not the case for
the present sample of galaxies which show sometimes low or even absent HeII/\Hb~ line ratios (see Paper I).
The modelling of these line ratios is  problematic because the behaviour of He lines
as coolants of the gas throughout the clouds is less important due to their low abundance,  
at least by  factors of 10,  and can change the relative extent of  the  He$^+$ ion
domains throughout the emitting clouds (Figs. 2 and 3). Therefore, the  HeII/\Hb~  line ratios are  often unpredictable (Umeda, 2022).

\subsection{Classification of the sample galaxies}

\begin{table}
\centering
\caption{Classification of the sample galaxies}
\begin{tabular}{lccccccccccccc} \hline  \hline
\ galaxy & type  & model    & \Hb$_{calc}$$^1$ \\ \hline
\ N1       & AGN   & ejec  & 0.005                \\
\ N3    & AGN   & accr     & 1.8                  \\
\ N5     &  SB   & accr     & 0.09                 \\
\ N6     &AGN    & accr     &0.06                  \\
\ N7     &SB     &ejec      & 0.5                  \\
\ N8a    &AGN    &ejec     & 1.75                  \\
\ N8b    &SB     &accr     &0.156                  \\
\ N9     &AGN    &ejec     &0.37                   \\
\ N10a   &AGN    &accr      &0.43                  \\
\ N10b   &SB     &ejec     &6.2                    \\
\ N11a   &AGN    &accr     &1.33                   \\
\ N11b   &SB     &accr     &0.13                   \\
\ N12a   &AGN    & accr    &0.71                   \\
\ N12b   &SB     & ejec    &7.1                    \\
\ N13a   &AGN    &accr     &0.34                   \\
\ N13b   &AGN    &ejec     &0.37                   \\
\ N13c   &SB     &ejec     &4.0                     \\ \hline

\end{tabular}

$^1$ in \erg, calculated at the nebula (Tables A.3-A.8)

\end{table}

In Table 2 we present the classification of the spectra which results from our modelling method. 
The data were  provided by Nakajima et al. 
The models were selected adopting AGN, SB, accretion and ejection models.  
For all the SDSS galaxies (N10-N13) and for HSC N8 we found that both an AGN and an SB model are  valid and can contribute to the spectrum.  
The galaxies showing a double or 
even a higher number of  radiation sources are generally the product of merging (Contini 2012, 2013). 
The types selected by our models  are only a first trial  because,  
whichever method is adopted by the calculations,  it  leads to approximated results. The main cause  is due  
to the  adopted coefficient exactness and also to  the   observed line types in each spectrum which do not cover 
the whole ionization stages and therefore do not  strongly constrain the models. 

\noindent
Tables A.3-A.8 and Figs. 4-7  suggest that:

N1 is an ejecting  AGN because [OIII]4363/\Hb~ and [OII]/\Hb~  ratios are best fitting the data.  
N1  SB  ejection model  should be eliminated because [OIII]4363/\Hb~ is  underpredicted by an error of 29 \% 
and HeII/\Hb~   reproduces the observed  upper limit instead  of  underpredicting it. 
The SB accretion model for [OIII] 5007+/\Hb~ corresponds to an error 
of 5.6 \% while the error for AGN ejection is 1.5 \%.  Then, AGN ejection is selected for the N1 galaxy but SB 
accretion could contribute to some of the lines. 

N3 is an accreting AGN because ejection has an [OIII]5007+/\Hb~	 error of 28\% while accretion of 4.6\%  and
an error of 2.5 \% for [OIII]4363/\Hb~ accretion and of 5\% for [OIII]4363/\Hb~ ejection.
The best fitting SB model has an [OIII]4363/\Hb~ error
of  25\% for accretion  and of 30 \% for ejection, an error of 5.5\% for [OIII]5007+/\Hb~ accretion and of 1.5\% for 
 ejection.
An  error of 100\% for HeII/\Hb~ SB ejection eliminates this model. An error of 43\% for [NII]/\Hb~ for SB accretion is 
also rather high.  Therefore we tend to select the AGN accretion model for N3.
 
N5 AGN accretion shows an error of 17.5\% for [OIII]5007+/\Hb~ and of 6.3\% for ejection, while N5 SB 2.2\% and 6.9 \%,
respectively.  HeII/\Hb~ eliminates  AGN ejection because it overpredicts the upper limit. 
The error of  SB [OII]/\Hb~  is  28\% for ejection while of 3.7 \% for accretion eliminating N5 SB ejection.
The errors for  AGN [OII]/\Hb~ are within 4.7. Therefore we select  SB accretion for N5.

N6 AGN  [OIII]4363/\Hb~ shows errors of 10 \% and 20 \% for accretion and ejection, respectively. 
All the other lines  well fit the data.
For SB models  the [OIII]4363/\Hb~  ratios have errors of 48 \% and 27 \% , respectively. The worst fit is for SB accretion 
HeII/\Hb~  with an error of 93 \%.
Therefore we select the AGN accretion model to represent N6 and we less recommend  the SB ejection model which shows an  
error of 10\% for the [OIII]5007+/\Hb~ line ratio.

N7 spectrum constrains the models because [OIII]4363/\Hb~  is an upper limit,
therefore the only fitting model is for an ejecting  SB (see Fig. 4).

N8 can be selected as an ejecting AGN although the error of 12.5\% for [OII]/\Hb~ ejection is higher than 
for accretion (10\%) but [OIII]5007+  reproduces better the observed line for ejection. 
The eventual SB model suggests accretion. 

N9 spectrum has been reproduced  by good precision  for the AGN  [OIII]5007+/\Hb~ and [OIII]4363/\Hb~  ratios  in the ejection case, however,
the \Ha/\Hb~ ratio has an error of 56 \%. Therefore the accretion model is selected. From this model we obtain that the [NII]/[OII] ratio
is log(0.047)=-1.33 and we can find the [NII]/\Hb~ true value neglecting the upper limit of  0.009. 
As for the SB models, we hardly select accretion due to a 23\% error for [OIII]5007+/\Hb~ and of 98\% for HeII.

N10 seems definitively an accreting AGN. However an ejecting SB can be accepted as a secondary 
component  with an error of 13.6 \% for [OII]/\Hb~ and of 18\% for [OIII]4363/\Hb.
AGN and  SB can coexist in the same  galaxy which  results from merging. The SB does not need to be central 
and an AGN can be displaced from the central zone  (Contini 2012).

N11  spectrum is well reproduced by both AGN and SB accretion models even with an error of 10\% for the 
SB [OIII]5007+/\Hb~  and of 18\% for AGN [OIII]4363/\Hb.

N12 spectrum is well reproduced by an accreting AGN with an error of 13.6\% for [OIII]4363/\Hb~ and by
an ejecting SB with an error of
13.6\% for [OII]/\Hb~ which is as weak as  [OIII]4363/\Hb. Both can be adopted for N12 which is perhaps a merging product.

N13 spectrum is well  fitted by both accretion and ejection AGN models  and by an ejecting SB because the 
accretion [OIII]5007+/\Hb~  line ratio shows an error of 13.8 \%.

\begin{figure*}
\centering
\includegraphics[width=6.6cm]{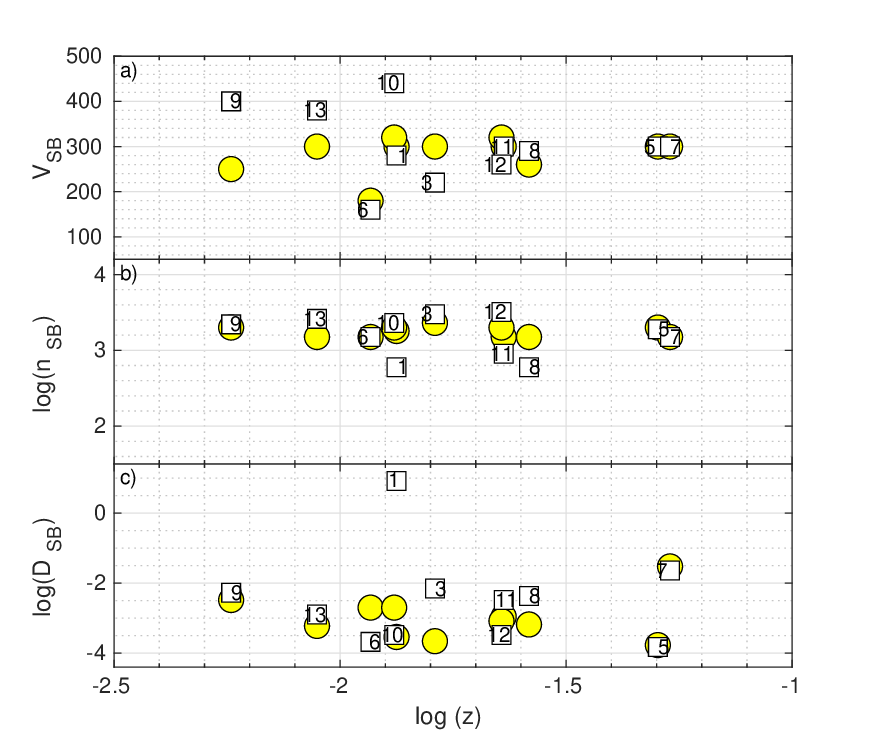}
\includegraphics[width=6.6cm]{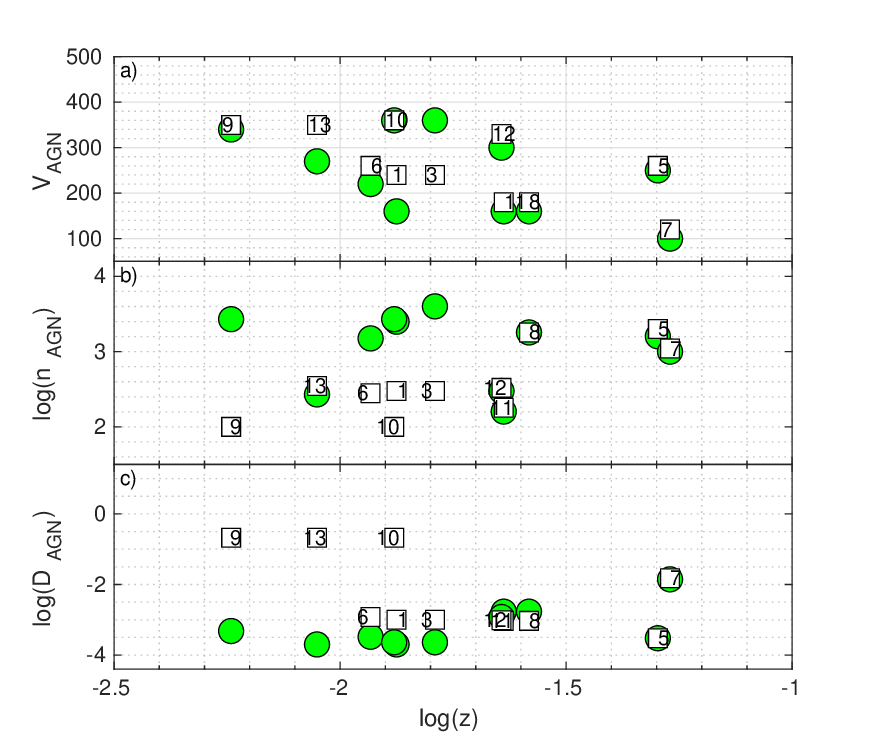}
\includegraphics[width=6.6cm]{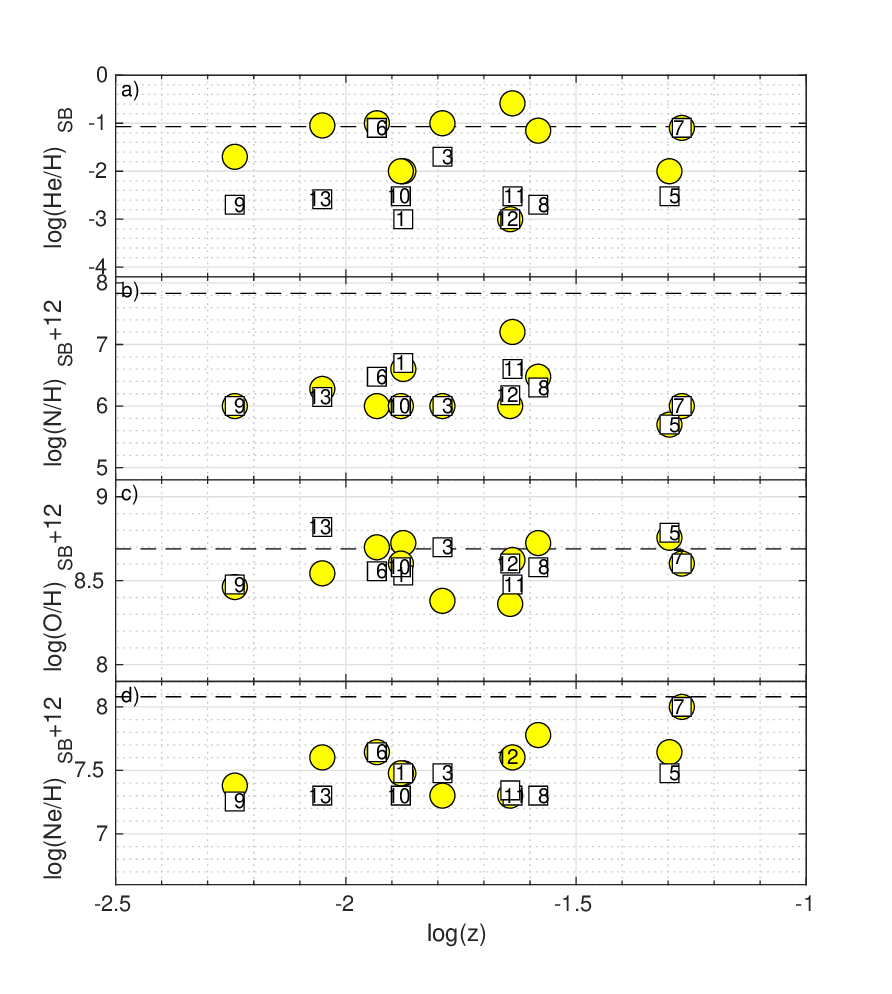}
\includegraphics[width=6.6cm]{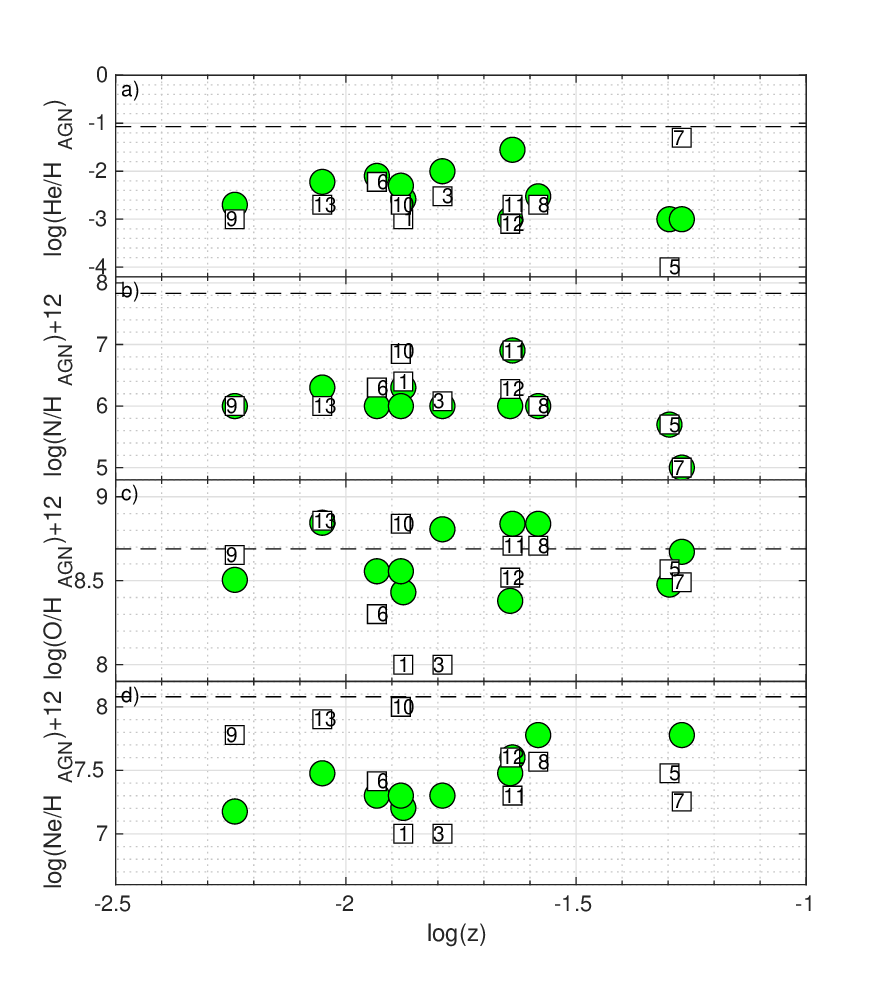}
\includegraphics[width=6.6cm]{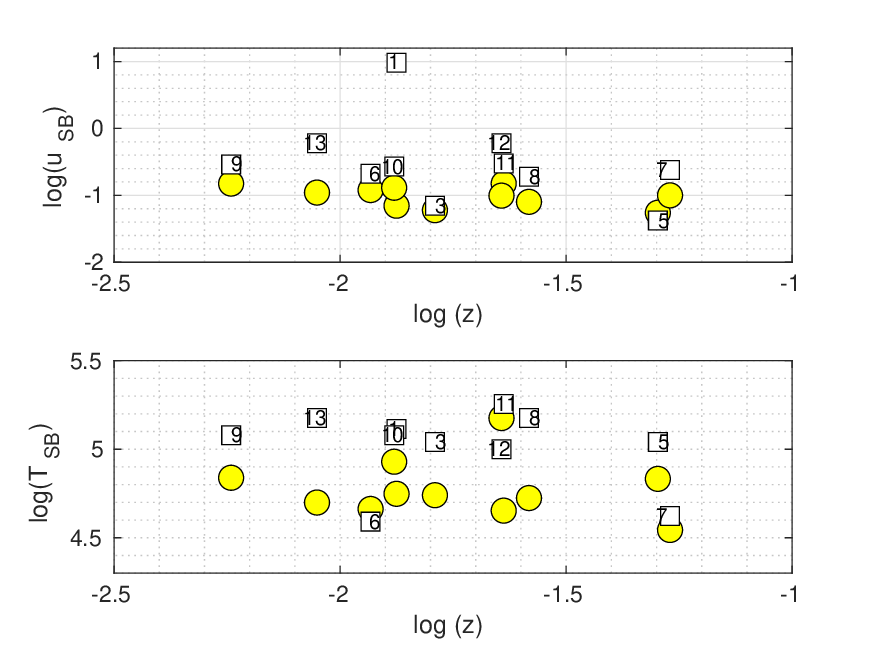}
\includegraphics[width=6.6cm]{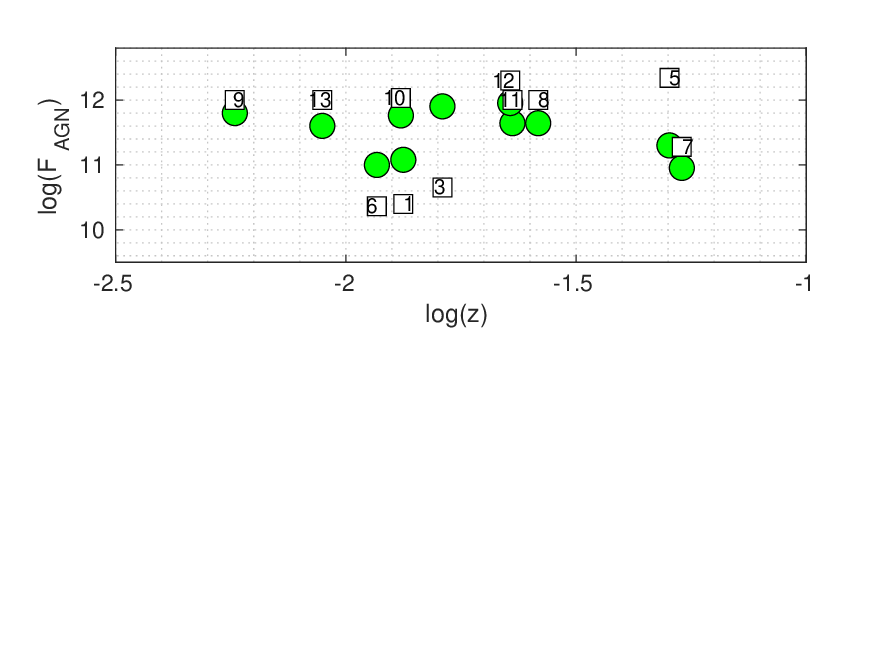}
\caption{Left. Distribution of the calculated  physical parameters representing the shock (top panels), 
of the calculated element relative abundances (middle panels) and of the radiation parameters (bottom diagrams) 
	as function of z for  SB (left) and AGN models (right). 
Yellow circles: accretion, white squares: ejection.
Right. The same for AGN models but green is for accretion and white for ejection.
Dashed lines show the solar abundance ratios to H (Grevesse  2019).
	}

\label{inpz}

\end{figure*}

\subsection{Calculated parameter trends with z}

Nakajima et al   assumed SB  models for all the galaxies.
To complete the investigation we report in Fig. 8   the results for  the sample objects, adopting 
both AGN and SB accreting and ejecting models. They are presented as function of the redshift.  
The calculated parameters were  selected   
by  reproducing the data  within  errors $<$ 20\% for the strong lines and  $<$50\% for the weak ones.  
Our aim is  to compare O/H metallicity with Nakajima et al results and to complete the modelling for each  galaxy.
In Fig. 8  (panels at the left) the yellow circles refer to SB accretion  and the  white squares to  SB ejection.
In the  panels at the right the  green circles refer to AGN accretion  and  white squares to  AGN ejection.
The redshift range is relatively small and it is included within the local universe. Therefore we
wonder if  extrapolating towards higher  z could make sense. The number of the sample galaxies
is also small, anyhow some interesting trends  can be noticed.

Fig. 8 top  panel is  dedicated  to the  shock velocities (\Vs ~ in \kms), 
the preshock densities
(\n0 ~ in \cm3) and to the cloud geometrical thickness ($D$ in pc). 
Only  \Vs ~ are reported on a linear scale (not logarithmic as for the other parameters) because they range
 between 100 \kms and 500 \kms. Moreover, they are always referred to in  these unities.
Shock velocities  are similar to those in the NLR of AGN.
Notice the decreasing trend of \Vs ~ in AGN models with increasing z
and  the  increasing  \n0 in agreement with the Rankine-Hugoniot law for  the conservation of  mass. 
This evidence  suggests that the shock velocity is a prime parameter in the interpretation of AGN spectra.
However, this  result characterizes only the  AGN models because
it seems that   \Vs~  in the  SB model trend more likely increases with z while \n0  remains constant.
 The pre-shock densities are low in the  AGN ejection case  because the jets collide with the ISM clouds.
 Ejection models for N9, N10 and N13 characterized by the highest \Vs ~ overcome the other galaxies 
 not only  regarding both AGN and SB \Vs ~  but also for
 the AGN geometrical thickness of the ejecting clouds and for the AGN O/H and Ne/H relative abundances
 (middle diagrams). 
The results show that the clouds are
highly fragmented by turbulence near the shock front for both SB and AGN. $D$ ranges within
10$^{-4}$ pc and 10$^{-3}$pc for AGN, except for N9, N10 and N13 with $D$$>$0.1 pc.
For SB models the $D$ range is  similar (10$^{-2}$-10$^{-4}$pc) except for galaxy N1 with $D$$\sim$3 pc.  
N1 also corresponds to an  overwhelming
ionization parameter $U$ (left bottom panel) which  keeps the cloud  gas ionized up to a large distance from 
the impinging  edge. 

The photoionizing flux and star temperatures  are also prevailing  in N9, N10 and N13 ejection models
 for SB galaxies (left bottom panel). 
 Also  the AC flux in AGN ejection models (right bottom diagram) 
 is slightly higher than that calculated  by accretion for the same galaxies. 
 These arguments are valid only  for galaxies at redshift $<$0.013. 
These parameters are closer to those calculated for AGN at higher z.
In the bottom left  panel SB models clearly show that $U$ and T are higher for ejection 
than for accretion and both slightly decrease towards higher z.
Except for N6  which  shows a very low temperature (bottom diagram) the ejecting clouds within   
SB galaxies  are  warmer than the accreting ones. It seems that the ejected matter is  heated by a 
higher ionizing radiation  flux which derives
from stars close to outbursts while in the accretion case the stars are in a more quiescent phase. 

In the middle left  panel  the relatively high He/H ratio for most of the SB accretion models 
stands out because  close to solar. The same occurs for a few accretion models which show
O/H solar abundances for  SB (N6, N10, N8 and N5). N12 and N7 are approximated.
AGN models overcome the O/H solar abundance for both accretion and ejection in N13, in N10 ejection  and
in N3, N11 and N8 accretion.

\begin{table}
\centering
\caption{Symbols in Figs.  11 and 12 }
\begin{tabular}{llcl} \hline  \hline
\ symbol   &object                       & Ref. \\ \hline
\  encircled dot&  SLSNR hosts           & (1) \\
\  triangles at z$\geq$0.1 & SLSNII hosts & (2)\\
\  filled triangles & SLSNI hosts         & (3)\\
\  square+dot & only shock models for SLSN hosts & (4)\\
\  square+cross& Type Ic host central     & (5)\\
\  circle+cross & Type Ic host at SN positions & (6) \\
\  triangles  at z$\leq$ 0.1 & SN Ib host & (7)    \\
\  encircled triangles  & SN IIb hosts    & (8) \\
\  opposite triangles    & SN Ic hosts    & (9) \\
\  double triangles      & SN IcBL hosts  & (10) \\
\  hexagrams               & SN Ibc        & (11)  \\
\  asterisks &GRB hosts                   & (12) \\
\  triangle+cross    & LGRB hosts         & (13)  \\
\  pentagrams  & LGRB different hosts     &  (14)  \\
\  triangle +plus& LGRB hosts with WR stars& (15) \\
\  encircled asterisks &LGRB at low z     & (16)\\
\   hexagrams  & SGRB hosts        & (17)\\
\  dots & starburst galaxies              & (19,25)\\
\  green open circles & AGN              & (20,21,25)  \\
\  filled circles & LINER                 & (22) \\
\  plus &  low-luminosity nearby galaxies & (23)  \\
\  cross& HII regions in local galaxies   & (24)  \\
\  green open circles& AGN                & (27)  \\ \hline
\  
\end{tabular}

 (1), (2), (3), (4) (Leloudas et al 2015);
(5), (6) (Modjaz et al 2008);
(7), (8), (9), (10), (11) (Sanders et al 2012);
(12) (Kr\"{u}hler et al 2015);
(13) (Savaglio et al 2009);
(14) (Contini 2016a, table 8);
(15) (Han et al 2010);
(16) (Niino et al 2016);
(17) (de Ugarte Postigo et al 2014);
(19), (20) (Contini 2014);
(21) (Koski 1978, Cohen 1983, Kraemer et al 1994, Dopita et al 2015);
(22) (Contini 1997);
(23) (Marino et al 2013);
(24) (Berg et al 2012);
(25) (Contini 2016b);
(27) (Dors et al 2021);

This table is updated from Contini (2017, Table 9).

\end{table}

\begin{figure}
\centering
\includegraphics[width=8.8cm]{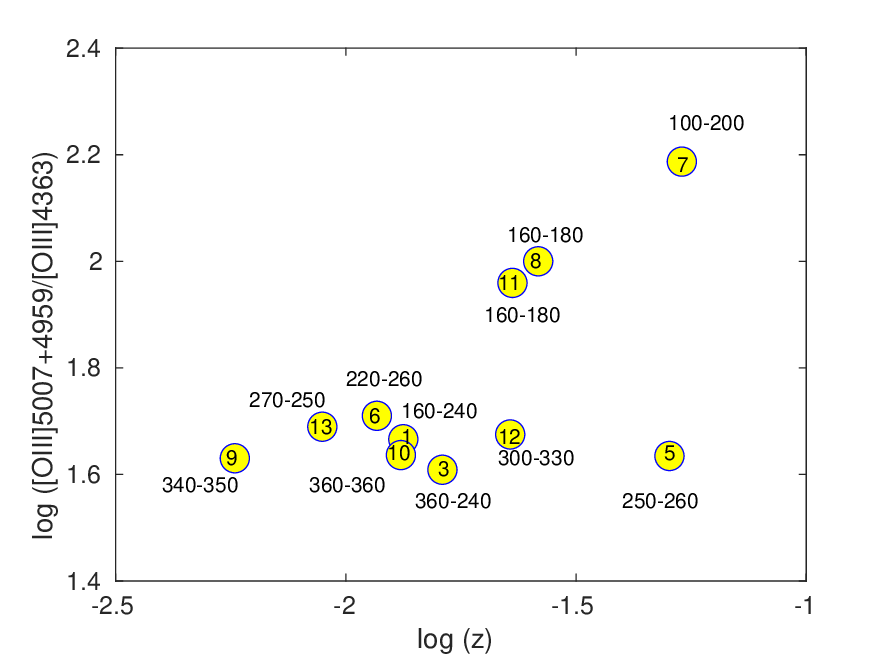}
\caption{Observed [OIII]5007+4959/[OII]4363 vs redshift (yellow filled circles). Numbers refer to the shock 
velocities calculated by the models in \kms  presented in Appendix A}

%\label{RO3}

\end{figure}

\begin{figure*}
\centering
\includegraphics[width=6.8cm]{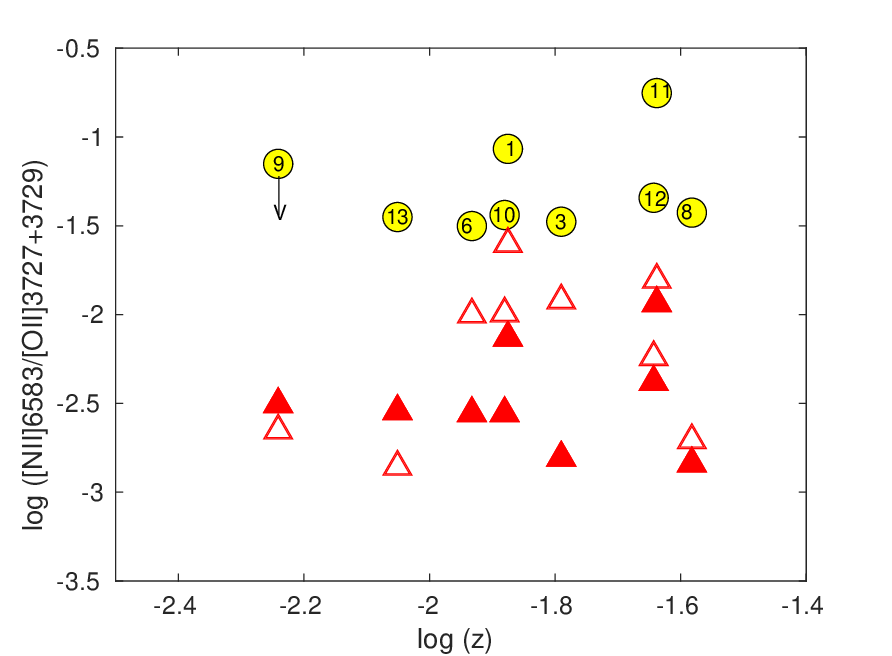}
\includegraphics[width=6.8cm]{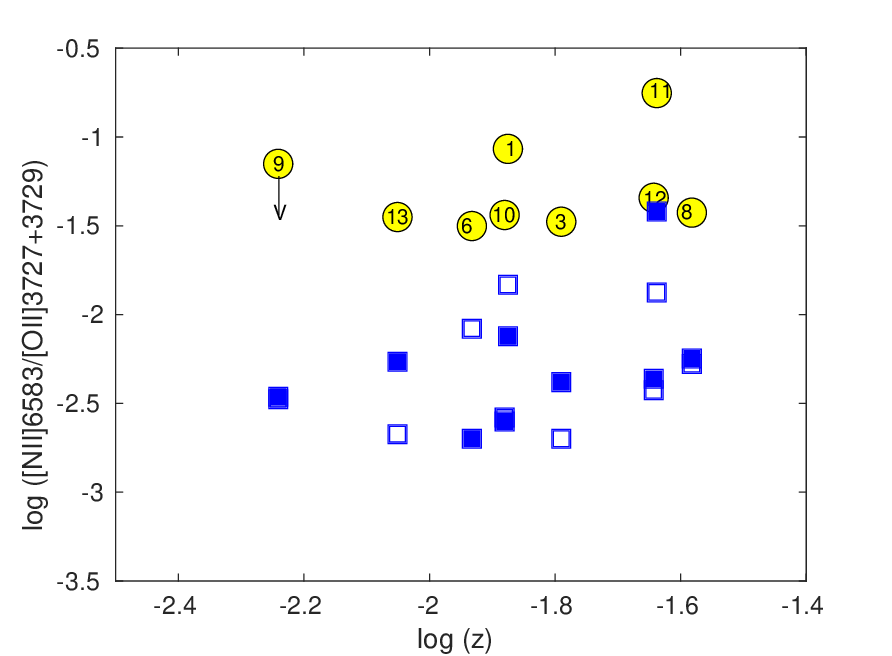}
\caption{Observed [NII]6583/[OII]3727+3729 vs  redshift (yellow filled circles).
 We have added in Fig. 10 left panel the calculated  log(N/O)
as function of z for AGN models (red triangles) and in the right panel the  calculated log(N/O) as function of z for
SB models (blue squares). 
Filled is for accretion and open  is for ejection.
The Y-axis scale is the same for all the ratios. (see sect. 3.4.2, first paragraph).
}

\end{figure*}

\begin{figure*}
\centering
\includegraphics[width=10.8cm]{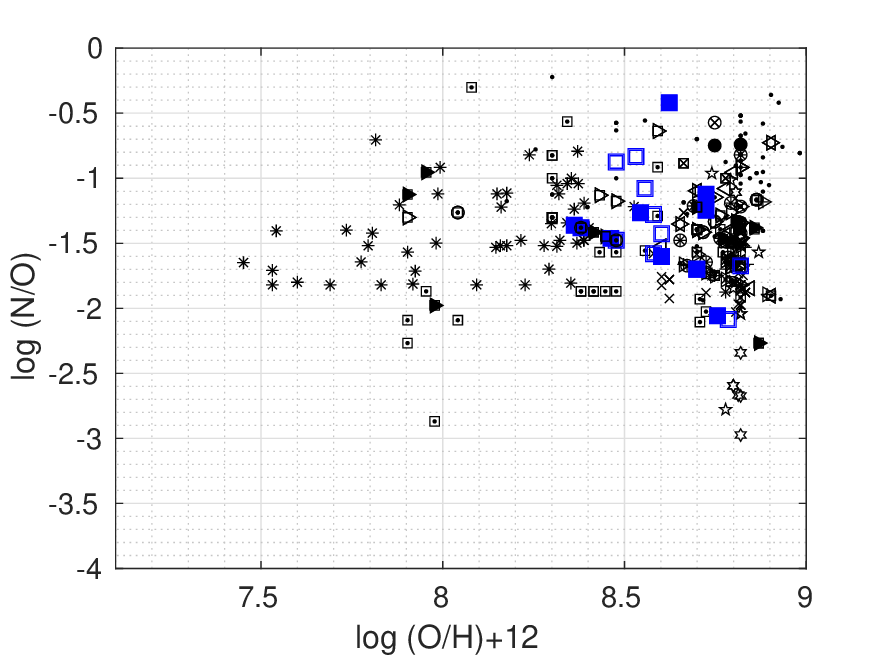}
\caption{
 N/O relative abundances as function of 12+log(O/H) calculated in this paper  by 
	  SB models (blue squares).   Open is for ejection,  filled for accretion.
	The calculated N/O ratios are shifted towards higher values by a factor of 10.
Symbols for the  other galaxy types  are  explained in Table 3.
}
%\end{figure*}

%\begin{figure*}
\centering
\includegraphics[width=10.8cm]{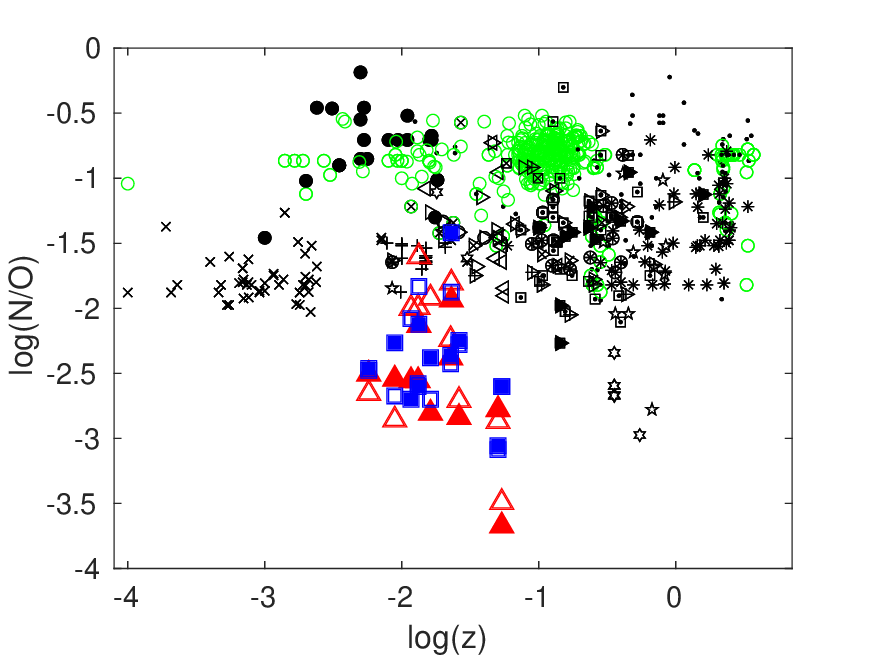}
\caption{
N/O relative abundances   as  function of log(z) calculated in this paper.
Symbols as in Fig. 11.
}
\end{figure*}

\begin{table*}
\centering
\caption{Comparison of oxygen metallicity}
\begin{tabular}{lccccccccccccc} \hline  \hline
\ galaxy            & 12+log(O/H)$^1$ &12+log(O/H)$_0$$^2$ & 12+log(O/H)$_1$$^3$&12+log(O/H)$_0$$^4$ & 12+log(O/H)$_1$$^5$\\ \hline
\ HSC  J0845+0131 (N1)  & 7.35  &8.43 &8.0  &8.72    & 8  \\
\ HSC  J0935-0115 (N3)  & 7.17  & 8.8 &8.0  &8.38   & 8.7 \\
\ HSC  J1237-0016 (N5)  & 7.55  &8.48 &8.57 &8.75    & 8.56\\
\ HSC  J1401-0040 (N6)  & 7.64  & 8.55&8.3  &8.7    & 8.3\\
\ HSC  J1407-0047 (N7)  & --    &8.67 &8.49 &8.6    &8.49  \\
\ HSC  J1411-0032 (N8)  & 8.12  &8.84 &8.7  &8.72    & 8.7\\
\ HSC  J1452+0241 (N9)  & 7.2   &8.5  &8.65 &8.46    & 8.65\\
\ SDSS J1044+0353 (N10)  & 7.45 &8.55 &8.84 &8.6    & 8.84\\
\ SDSS J1253-0312 (N11)  &8.09  &8.84 &8.7  &8.62    & 8.16\\
\ SDSS J1323-0312 (N12)  &7.74  &8.38 &8.52 &8.36    & 8.5\\
\ SDSS J1418+2102 (N13)  & 7.64 &8.84 &8.86 &8.54    & 8.36 \\ \hline
\end{tabular}

$^1$ calculated by Nakajima et al.;
$^2$ calculated by models mh1-mh13 (str=0);
$^3$ calculated by models mh1-mh13 (str=1);
$^4$ calculated by models ms1-ms13 (str=0);
$^5$ calculated by models ms1-ms13 (str=1)

\label{sum}

\end{table*}

\subsection{Significant line ratios}  
\subsubsection{\RO3}  

In Fig. 9 the observed  [OIII]5007+4959/[OIII]4363 (=\RO3) line ratios are  reported as function of the redshift
for the sample galaxies.
Special attention is given to  \RO3~  because we  learned during the
spectral fitting process that the [OIII]4363 /\Hb~ line ratio constrains  significantly  the models.
\RO3 does not directly depend on the O/H relative abundance, therefore we shall look for
the physical parameters responsible for the different trends.
Most of the galaxies show \RO3~ between 40 and 50.
N7 (upper limit), N8 and N11 are outstanding.  All of them  correspond to [OII]/\Hb $\sim$0.8.
Then we can include in this   series also N6.
It is clear that these galaxies   show  an increasing trend of \RO3 with z,
different from the other galaxies of the sample which  are nearly constant.
The galaxy series are  distributed following the shock velocities which determine the temperature of the
gas downstream. At higher velocities the  temperature of the gas emitting the  [OIII]4363 line strongly  
increases reducing the \RO3 ratio.
The scattering of the data shows that also the gas density and the photoionizing flux intensity are acting.
Another series of galaxies  (N5, N12 and N13) in the present sample includes  those characterized
by a strong flux $\geq$ 10$^{12}$  photons cm$^{-2}$ sec$^{-1}$ eV$^{-1}$ at the Lyman limit, in the ejection case.

\subsubsection{[NII]/[OII]}  

In Fig. 10  [NII]/[OII] line ratios are shown as function of the redshift.
[NII] and [OII] are linked by charge exchange reactions
therefore  increasing or decreasing  trends  depend at most
on the N/O relative abundance. N5 and N7 do not appear in the diagrams because  no data for [NII] were reported.
N9 is an upper limit.
We have added in Figs. 10  the N/O abundance ratios for each galaxy calculated by SB (right) and AGN models (left), 
respectively. 
 Line intensity ratios  are calculated by multiplying one term which depends  on the collision strengths
(and  derives from quantum mechanics),  another term which depends on  thermal equilibrium and  a term which  contains
 the  relative abundances  of the elements. If charge exchange reactions link the two lines, the result  will be proportional
in particular to  the relative abundance of the elements. Therefore the [NII]/[OII] trend will follow the N/O trend. This can be roughly
seen in Figure 10 panels, considering  that for  galaxy N9 the line ratio is an upper limit.
To recover their values N/O ratios should be multiplied by the factor  which depends on the collisional parameters
of both lines and  on the temperature of the emitting gas (Williams 1967).
 The sample is  poor in number of galaxies and it covers a restricted redshift range  which is 
 less adapted to extrapolate to higher z. Alternatively,  the different trends could be interpreted 
 as simple fluctuations.

 \subsubsection{N/O abundance ratios and bursting-stars formation}

Figs. 11 and 12 are updated from Contini (2017, Figs. 6 and 7, respectively). In Fig. 11 we  compare log(N/O) as function of 12+log(O/H) 
for different SB galaxies  (see Table 3, updated from Contini 2017, Table 9) accounting in particular  for the present model results 
 for Nakajima et al EMPGs  and (Fig. 12) also for  the rich sample of AGN presented by Dors et al (2021).

Although the Nakajima et al sample of EMPGs is not rich in number of objects,   we  have investigated the log(N/O) trend as function of 12+log(O/H)
of SB dominated galaxy model results in Fig. 11.
First we  have shifted our results multiplying N/O  (large blue squares, filled for accretion and open for ejection) by a factor of 10.
In this way  EMPGs cover the region between GRB host  galaxies and super-luminous SN hosts.
Henry, Edmunds \& Koppen (2000) claim that intermediate-mass stars between 4 and 8 \msol dominate
nitrogen  production while massive stars dominate oxygen production. Nitrogen is primary at low metallicity but when 12+log(O/H)$>$8.3
secondary N becomes prominent.
At 12+log(O/H) $\sim$8.25 log(N/O)  turns upwards and rises steeply. For Mouhcine \& Contini (2002) it turns upwards steeply at 8.7,
similarly to our results for EMPGs which turn up steeply at $\sim$ 8.7.
This happens because nitrogen synthesis in intermediate mass stars is increasing with metallicity while oxygen production in massive stars
is decreasing.
However, at 12+log(O/H) $>$ 8.3  log(N/O) calculated by our models for EMPGs slightly decreases  towards 12+log(O/H)=8.7,
opposite to Henry et al fig.1b
and Mouhcine \& Contini Fig. 1 trends. Mouhcine \& Contini,  in particular,  claim that "for
bursting-star formation, the observed dispersion in the N/O vs O/H relation is explained by the time delay between the release of oxygen by
massive stars  into the ISM and that of nitrogen by intermediate-mass stars. During the starburst event, as massive stars dominate
the chemical enrichment, the galaxy moves towards the lower right part of the diagram in their Fig. 1. After the end of the burst
nitrogen increases at constant oxygen."

Fig. 11 shows that N/O ratios calculated for EMPGs are lower  by a factor of $\sim$ 10 than those calculated  for  other galaxies
with the same metallicity.
 Mouhcine et  \& Contini suggest that "the low N/O abundance ratio with respect to other  galaxies with comparable metallicities is well explained
 if they have just undergone a powerful starburst
 which have enriched their ISM in oxygen. In these objects nitrogen may have not been completely released."

We would like to explore how the N/O  trend   develops  with the redshift. In  Fig. 12 the log(N/O)
ratio calculated for the Nakajima et al sample is compared with  different galaxy types
described  in Table 3.  As well as for Fig. 11, they  were selected on the basis that the parameters 
were calculated by the same code.  At present we can roughly recognise 
a peak for LINERs at z$\sim$0.006,  a rather poor  peak for  SB galaxies  at z$\sim$1.0 
 and a double  maximum for  SB accretion and AGN ejection models at z $\sim$0.012  for EMPG models.
 Fig. 12 confirms  that EMPGs are  extremely nitrogen-poor objects.

\section{Concluding remarks}

The  galaxy sample presented by   Nakajima et al. (2022)  between z=0.00574 and z=0.05045  
 was selected  because of  oxygen deficiency.
We have modelled the spectra by  the {\sc suma} code which accounts for both photoionization and shocks because
it can be used to  distinguish between accretion towards and ejection of matter from the radiation source.
The detailed analysis  of the line ratios presented in the first sections shows that the EMPGs in the Nakajima et al sample
have   various element  abundances  and  morphologies. 
 We have adopted in this paper the updated solar log(O/H)+12=8.69 by Grevesse (2019). Then, in Fig. 11 a few  SB dominated EMPGs show 
near-solar O/H abundance ratios, and a few AGN dominated models show O/H  even higher than solar (see Fig.8), which is in agreement with 
AGN in general.
Our results demonstrate that low oxygen line  ratios to \Hb~ not always indicate a proportional  O/H relative  deficiency,
such as to justify their classification as EMPG. Table 4  shows that the oxygen metallicities calculated by our models are higher 
than those calculated by the strong line method.
On the other hand N/H ratios  reach  values lower  than  (N/H)$_{\odot}$ by factors $\sim135$.  
Therefore, our results justify the classification of the Nakajima et al sample among EMPG  if nitrogen is also included among the "extremely
poor metals".  He/H  are depleted by factors 
between 10 and 100 when AGN models are adopted as well as for SB models
 in the ejection case. SB  accreting models show  about solar  He/H abundances ratios which indicate
 that accreted matter is coming from the ISM.
We suggest a  classification of the Nakajima et al sample objects  selecting the models  which best 
reproduce the spectra. 
All the SDSS objects and HSC N8 galaxy can be interpreted as merging products on the basis 
of a composite AGN + SB nature.

In the second  part of the  paper, although the Nakajima et al sample of EMPGs is not rich in number of objects,  
we  have  discussed the log(N/O) trend as function of 12+log(O/H) and of the redshift.
Bursting-star trends  are confirmed  as well as  strong metal deficiency, in particular, of nitrogen. 
The N/O trend as function of the redshift  gives only a hint for the AGN-LINER
and SB peak locations in general.

%\begin{acknowledgements}
%\end{acknowledgements}	

{}

\newpage

\begin{appendix}

\section{Tables}

\begin{table*}
\centering
\caption{ Comparison of observed with calculated line ratios to \Hb=1 for the Nakajima et al. galaxies}
\begin{tabular}{llccccccccccccc} \hline  \hline
\   HSC      &N1    &mp1.0 &mp1.1&N3    & mp3.0&mp3.1& N5   &mp5.0&mp5.1&N6   &mp6.0&mp6.1\\ \hline
\ [OII]3727+ &0.42 &0.36   &0.37 &0.21  & 0.23 &0.26  & 0.43 &0.39 &0.42&0.825 &0.72 &0.7\\
\ [NeIII]3896+&0.29&0.31   &0.31 &0.25  &0.3   &0.27  & 0.71 &0.75 &0.84&0.42  &0.34 &0.4\\
\ \Hg 4340    &0.443&0.46  &0.44 &0.445 &0.46  &0.45  & 0.49 &0.46 &0.46&0.5   &0.46 &0.45\\       
\ [OIII]4363  &0.085&0.07  &0.05 &0.08  &0.05  &0.036 & 0.167&0.125&0.08&0.124 &0.083&0.45\\
\ HeII 4686   &$<$0.01&0.9&0.09 &0.024 &0.06  &0.18  &$<$0.0128&0.19&0.22&0.03 &0.12 &0.12\\
\ \Hb 4861    &1    &1     &1    & 1    &1     &1     & 1    &1    &1   &1     &1    &1   \\
\ [OIII]5007+ &3.94 &4.0   &4.3  &3.25  & 3.25 &3.0   &7.2   &7.5  &7.0 &6.36  &7.0  &5.9\\
\ HeI 5876    &0.21 &0.21  &0.12 &0.086 &0.13  &0.1   &0.13  &0.11 &0.126&0.094&0.1  &0.08\\ 
\ \Ha 6563    &2.67 &2.9   &3.6  &2.53  &2.9   &3.16  & 2.84 &2.9  &3.0  &2.87 &2.9  &3.15\\
\ [NII]6583   &0.036&0.037 &0.03&0.007  &0.007 &0.009 &-     &0.007&0.007&0.026&0.04 &0.04\\ 
\ [SII]6716   &0.04  &0.03 &0.016&0.023 & 0.02 &0.016 &0.032 &0.024&0.18 &0.089&0.044&0.24 \\
\ [SII]6730   &0.02  &0.07 &0.033&0.01  & 0.04 &0.036 &0.03  &0.054&0.39 &0.06 &0.1  &0.49\\
\ \Hb$_{calc}$ $^0$ &-&6.7 &4.24 &-     &7.7   &3.75  &-     &2.3  &2.2  &-    &1.6  &1.5 \\
\ \Vs $^1$    &-     &200  &200  &  -   & 300  &300   &-     &250  &300  &-    &180  &180\\
\ \n0 $^2$    &-     &2000 &2000 &  -   & 2000 &2000  &-     &2000 &2000 &-    &1800 &1800\\
\  $D$ $^3$   &-     &0.0176&0.83&-     & 0.007&0.007 &-     &0.0037&0.0025&-  &0.007&0.007\\
\ $F$ $^4$    &-     &9e11 &9e11 & -    & 9e11 &8e11  &-     &8.6e11&9e11&-    &4.2e11&4.2e11\\
\ He/H        &-     &0.16 &0.1  &  -   & 0.1  &0.1   &-     &0.09&0.09  &-    &0.08  &0.08 \\
\ N/H $^5$    &-     &0.04 &0.04 &  -   & 0.01 &0.01  &-     &0.005&0.005&-    &0.03  &0.03\\
\ O/H $^5$    &-     &7.0  &7    &  -   & 6.6  &5.6   &-     &4.6  &5.6  &-    &7.8   &7.8\\
\ Ne/H $^5$   &-     &0.6  &0.5  &  -   &  0.6 &0.3   &-     &0.5  &0.5  &-    &0.4   &0.4\\
\ S/H $^5$    &-     &0.2  &0.02 &  -   & 0.2  &0.03  &-     &0.2  &0.2  &-    &0.2   &0.2\\
\ str         &-     &0    &1    & -    &0     &1     &-     &0    &1    &-    &0     &1   \\ \hline
\end{tabular}

$^0$ \erg; $^1$ \kms; $^2$ \cm3; $^3$ parsec; $^4$ photon cm$^{-2}$ s$^{-1}$ eV$^{-1}$ at the Lyman limit; $^5$ in 10$^{-4}$ units

\end{table*}

\begin{table*}
\centering
\caption{ Comparison of observed with calculated line ratios to \Hb=1 for the Nakajima et al. galaxies}
\begin{tabular}{llccccccccccccc} \hline  \hline
        \    HSC      &N7     &mp7.0&mp7.1&N8    &mp8.0 &mp8.1 & N9    &mp9.0&mp9.1\\ \hline
        \ [OII]3727+  &0.82  &0.74  &0.7   &0.8  &1.0    &1.0    &0.17  &0.23  &0.19 \\
        \ [NeIII]3896+&0.34  &0.3   &0.4   &0.74 &0.87   &1.0    &0.23  &0.27  &0.35 \\
        \ \Hg 4340    &0.47  &0.46  &0.45  &0.53 &0.46   &0.46   &0.46  &0.46  &0.45 \\
        \ [OIII]4363  &$<$0.026&0.06&0.037 &0.10 &0.13   &0.1    &0.078 &0.05  &0.026 \\
        \ HeII4686    &-     &0.12  &0.13  &-    &0.19   &0.23   &$<$0.013&0.068&0.813 \\
        \ \Hb 4861    &1     &1     &1     &1    &1      &1      &1       &1    &1    \\
        \ [OIII]5007+ &4.0   &4.6   &4.0   &10.0 &10.5   &11.9   &3.23    &3.24 &2.2  \\
        \ HeI 5876    &0.092 &0.11  &0.13  &0.106&0.12   &0.12   &0.13    &0.13 &0.11  \\
        \ \Ha 6563    &2.78  & 2.9  &3.12  &3.0  &2.9    &2.98   &2.83    &2.9  &3.1   \\
        \ [NII]6583   &-     &0.02  &0.02  &0.03 &0.02   &0.02   & $<$0.012&0.007&0.007\\
        \ [SII]6716   &0.094 &0.11  &0.096 &-    &0.06   &0.15   &0.016    &0.011&0.04   \\
        \ [SII]6730   &0.1   &0.24  &0.2   &0.05 &0.13   &0.25   &0.013    &0.02 &0.09  \\
        \ \Hb$_{calc}$ $^0$ &-&2.1  &1     &-    &1.27   &0.94   &-        &7.8  &4.4   \\
        \ \Vs $^1$    &-    &180    &210   &-    &160    &160    &-        &300  &300  \\
        \ \n0 $^2$    &-    &1500   &1500  &-    &1800   &1800   & -       &2000 &2000 \\
        \  $D$ $^3$   &-    &0.03   &0.013 &-    &0.0017 &0.0017 &-        &0.007&0.0077  \\
        \ $F$ $^4$    &-    &4.2e11 &4.2e11&-    &4.4e11 &4.4e11 &-        &9e11 &9e11  \\
        \ He/H        &-    &0.1    &0.1   &-    &0.1    &0.1    &-        &0.1     &0.1   \\
        \ N/H $^5$    &-    &0.01   &0.01  &-    &0.01   &0.01   &-        &0.01    &0.01  \\
        \ O/H $^5$    &-    &4.6    &4.6   &-    &6.9    &6.9    &-        &6.2     &5.2  \\
        \ Ne/H $^5$   &-    &0.3    &0.3   &-    &0.6    &0.6    &     -   &0.5     &0.5  \\
        \ S/H $^5$    &-    &0.2    &0.2   &-    &0.1    &0.09   &-        &0.1     &0.1  \\
        \ str         &-    &0      &1     &-    &0      &1      &-        &0       &1     \\ \hline
\end{tabular}

 $^0$ \erg; $^1$ \kms; $^2$ \cm3; $^3$ parsec; $^4$ photon cm$^{-2}$ s$^{-1}$ eV$^{-1}$ at the Lyman limit; $^5$ in 10$^{-4}$ units
\end{table*}

\begin{table*}
\centering
\caption{ Comparison of observed with calculated line ratios to \Hb=1 for the Nakajima et al. galaxies}
\begin{tabular}{llccccccccccccc} \hline  \hline
\   HSC      &N1    &mh1.0 &mh1.1&N3    & mh3.0&mh3.1 & N5   &mh5.0&mh5.1&N6   &mh6.0&mh6.1\\ \hline 
\ [OII]3727+ &0.42 &0.5    &0.47 &0.21  & 0.20 &0.21    & 0.43 &0.44 &0.45&0.825 &0.85 &0.83\\
\ [NeIII]3896+&0.29&0.32   &0.22 &0.25  &0.28  &0.2     & 0.71 &0.71 &0.73&0.42  &0.41 &0.45\\
\ \Hg 4340    &0.443&0.46  &0.46 &0.445 &0.46  &0.46    & 0.49 &0.46 &0.46&0.5   &0.464&0.46\\       
\ [OIII]4363  &0.085&0.10  &0.085&0.08  &0.082 &0.076   & 0.167&0.16 &0.18&0.124 &0.137&0.10 \\
\ HeII 4686   &$<$0.01&0.008&0.005&0.024 &0.027 &0.021  &$<$0.0128&0.004&0.018&0.03 &0.025&0.03\\
\ \Hb 4861    &1    &1     &1    & 1    &1     &1       & 1    &1    &1   &1     &1    &1   \\
\ [OIII]5007+ &3.94 &3.89  &4.0  &3.25  & 3.1  &4.17    &7.2   &8.46 &7.66&6.36  &6.0  &6.29\\
\ HeI 5876    &0.21 &0.003 &0.001 &0.086 &0.012 &0.02   &0.13  &0.001&0.001&0.094&0.009&0.005\\ 
\ \Ha 6563    &2.67 &2.92  &2.97 &2.53  &2.89   &2.95   & 2.9  &2.9  &3.2  &2.87 &2.92 &2.94\\
\ [NII]6583   &0.036&0.032 &0.04&0.007  &0.006 &0.007   &-     &0.007&0.007&0.026&0.021&0.027\\
\ [SII]6716   &0.04  &0.034&0.015&0.023 & 0.01 &0.012   &0.032 &0.02 &0.03 &0.089&0.031&0.03 \\
\ [SII]6730   &0.02  &0.079&0.03 &0.01  & 0.023&0.03    &0.03  &0.048&0.07 &0.06 &0.071&0.06 \\
\ \Hb$_{calc}$ $^0$ &-  &0.068&0.005&-     &1.8   &0.005   &-     &2    &3    &-    &0.05 &0.007\\
\ \Vs $^1$    &-     &160  &240  &  -   & 360  &240     &-     &250  &260  &-    &220  &260\\
\ \n0 $^2$    &-     &2500 &300  &  -   & 4000 &300     &-     &1600 &2000 &-    &1500 &280 \\
\  $D$ $^3$   &-     &2.e-4&1e-3&-     &2.3e-4&0.001    &-     &3.e-4 &3.e-4&-  &3.3e-4&1.2e-3\\
\ $F$ $^4$    &-     &1.2e11&2.5e10& -    & 8e11 &4.5e10 &-     &2.0e11&2.2e12&-  &1.0e11&2.3e10\\
\ He/H        &-     &0.0026&0.001&  -   & 0.01 &0.003   &-     &0.001&     &-  &0.008 &0.006\\
\ N/H $^5$    &-     &0.02 &0.025&  -   & 0.01 &0.012   &-     &0.007&     &-    &0.01  &0.02\\
\ O/H $^5$    &-     &2.7  &1    &  -   & 6.4  &1.0     &-     &3.0  &3.7  &-    &3.6   &2  \\
\ Ne/H $^5$   &-     &0.16 &0.1  &  -   &  0.2 &0.1     &-     &0.03 &0.3  &-    &0.2   &0.26\\
\ S/H $^5$    &-     &0.2  &0.03 &  -   & 0.2  &0.06    &-     &0.2  &0.15 &-    &0.15  &0.06\\
\ str         &-     &0    &1    & -    &0     &1       &-     &0    &1    &-    &0     &1   \\ \hline
\end{tabular}

$^0$ \erg; $^1$ \kms; $^2$ \cm3; $^3$ parsec; $^4$ photon cm$^{-2}$ s$^{-1}$ eV$^{-1}$ at the Lyman limit; $^5$ in 10$^{-4}$ units

\end{table*}

\begin{table*}
\centering
\caption{ Comparison of observed with calculated line ratios to \Hb=1 for the Nakajima et al. galaxies}
\begin{tabular}{llccccccccccccc} \hline  \hline
        \    HSC      &N7     &mh7.0&mh7.1&N8    &mh8.0 &mh8.1 & N9    &mh9.0&mh9.1\\ \hline 
        \ [OII]3727+  &0.82  &0.82  &0.84  &0.8  &0.88   &0.9    &0.17  &0.19  &0.18 \\
        \ [NeIII]3896+&0.34  &0.32  &0.45  &0.74 &0.74   &0.81   &0.23  &0.27  &0.26 \\
        \ \Hg 4340    &0.47  &0.46  &0.46  &0.53 &0.46   &0.46   &0.46  &0.46  &0.40 \\
        \ [OIII]4363  &$<$0.026&0.026&0.029 &0.10 &0.09   &0.09   &0.078 &0.07  &0.071 \\
        \ HeII4686    &-     &0.004 &0.16  &-    &0.018  &0.018  &$<$0.013&0.008&0.0085\\
        \ \Hb 4861    &1     &1     &1     &1    &1      &1      &1       &1    &1    \\
        \ [OIII]5007+ &4.0   &3.2   &3.6   &10.0 &9.5    &10.14  &3.33    &3.68 &3.5  \\
        \ HeI 5876    &0.092 &0.001 &0.074 &0.106&0.002  &0.001  &0.13    &2.1e-3&1.e-5 \\
        \ \Ha 6563    &2.78  & 2.9  &3.1   &3.0  &2.9    &3.14   &2.83    &2.9  &4.5   \\
        \ [NII]6583   &-     &0.015 &0.01  &0.03 &0.02   &0.02   & $<$0.012&0.009&0.002\\
        \ [SII]6716   &0.094 &0.06  &0.12  &-    &0.05   &0.04   &0.016    &0.01 &0.0016 \\
        \ [SII]6730   &0.1   &0.13  &0.23  &0.05 &0.1    &0.09   &0.013    &0.017&0.0023\\
        \ \Hb$_{calc}$ $^0$ &-&0.56 &0.51  &-    &1.36   &1.75   &-        &1.27 &0.37  \\
        \ \Vs $^1$    &-    &100    &200   &-    &160    &180    &-        &340  &350  \\
        \ \n0 $^2$    &-    &1000   &1100  &-    &1800   &1800   & -       &2700 &100  \\
	\  $D$ $^3$   &-    &0.014  &0.015 &-    &0.0017 &9.3e-4 &-        &4.8e-4&0.21  \\ 
        \ $F$ $^4$    &-    &9.0e10 &1.9e11&-    &4.4e11 &1.e12  &-        &6.3e11 &1e12  \\
        \ He/H        &-    &0.001  &0.05  &-    &0.003  &0.002  &-        &0.002   &0.001 \\
        \ N/H $^5$    &-    &0.01   &0.005 &-    &0.01   &0.01    &-        &0.01    &0.01  \\
        \ O/H $^5$    &-    &4.7    &3.1   &-    &6.9    &5.1    &-        &3.2     &4.5  \\
        \ Ne/H $^5$   &-    &0.6    &0.18  &-    &0.6    &0.37   &     -   &0.15    &0.6  \\
        \ S/H $^5$    &-    &0.1    &0.04  &-    &0.2    &0.1    &-        &0.1     &0.3  \\
        \ str         &-    &0      &1     &-    &0      &1      &-        &0       &1     \\ \hline
\end{tabular}

 $^0$ \erg; $^1$ \kms; $^2$ \cm3; $^3$ parsec; $^4$ photon cm$^{-2}$ s$^{-1}$ eV$^{-1}$ at the Lyman limit; $^5$ in 10$^{-4}$ units

\end{table*}

\begin{table*}
\centering
\caption{ Comparison of observed with calculated line ratios to \Hb=1 for the Nakajima et al. galaxies}
\begin{tabular}{llccccccccccccc} \hline  \hline
\  SDSS       & N10    &mh10.0 &mh10.1  &  N11& mh11.0& mh11.1&N12   &mh12.0&mh12.1&N13   &mh13.0&mh13.1 \\ \hline 
\ [OII]3727+  &0.22    &0.22   &0.28   &0.849 &0.90  &0.90  &0.176 &0.17 &0.17  & 0.48  &0.45  & 0.49\\
\ [NeIII]3896+&0.42   &0.44    &0.44   &0.52  &0.52  &0.5   &0.73  &0.89 &0.75  & 0.517 &0.51  &0.43\\
\ \Hg 4340    &0.48   &0.47    &0.40   &0.49  &0.464 &0.46  &0.49  &0.47 &0.47  & 0.48  &0.46  &0.40\\
\ [OIII]4363  &0.134  &0.133   &0.112  &0.11  &0.09  &0.093 &0.22  &0.19 &0.16  & 0.149 &0.145 &0.14\\
\ HeII 4686   &0.018  &0.02    &0.017  &0.017 &0.018 &0.016 &0.008 &0.008&0.008 & 0.022 &0.021 &0.017\\
\ \Hb 4861    &1      &1       &1      &1     &1     &1     &1     &1    &1     & 1     &1     &1\\
\ [OIII]5007+ &5.8    &5.9     &5.5    &10.02 &9.8   &10.6  &10.41 &10.88&11.6  & 7.29  &7.5   &7.2\\
\ HeI 5876    &0.086  &0.005   &2.3e-5 &0.11  &0.002 &0.001 &0.097 &6.e-4&1.6e-4 & 0.101 &0.007 &1.2e-5\\
\ \Ha 6563    &2.65   &2.89    &4.5    &2.74  &2.89  &3.14  &2.88  &2.87 &2.9   & 2.84  &2.9   &4.5\\
\ [NII]6583   &0.008  &0.008   &0.009  &0.15  &0.14  &0.18  &0.008 &0.009&0.009  &0.017 &0.018 &0.018\\
\ [SII]6716   &0.02   &0.005   &0.002  &0.072 &0.05  &0.04  &-     &9e-4 &6.4e-4 &0.046  &0.01  &2.9e-3\\
\ [SII]6730   &0.016  &0.012   &0.0026 &0.073 &0.1   &0.09  &-     &0.002 &0.0015&0.031 &0.03  &4.2e-3\\
\ \Hb$_{calc}$$^0$&-  &0.43    &0.39   &-     &1.33  &1.78  &-     &0.71  &1.55  &-     &0.34  &0.37\\
\ \Vs $^1$    &-     &360      &360    &-     &160   &180   &-     &300   &330   &-     &270   &350\\
\ \n0 $^2$    &-      &2700    &100    &-     &1800  &1800  &-     &1800  &1100  &-     &3000  &100  \\
\  $D$ $^3$   & -     &2.3e-4  &0.21   &-     &0.0017&9.3e-4&-     &0.0012&9.7e-4&-     &2.e-4 &0.21  \\
\ $F$ $^4$    &-      &5.8e11  &1.08e12&-     &4.4e11&1.e12 &-     &9e11  &2e12  &-     &4e11  &1e12\\
\ He/H        &-      &0.005   &0.002  &-     &0.028 &0.002 &-     &0.001 &0.0008& -    &0.006 &0.002\\
\ N/H $^5$    &-      &0.01    &0.07   &-     &0.08  &0.08  &-     &0.01  &0.019 &-     &0.02  &0.01 \\
\ O/H $^5$    &-      &3.6     &6.9    &-     &6.9   &5.1   &-     &2.4   &3.3   &-     &7.0   &7.2\\
\ Ne/H $^5$   &-      &0.2     &1.0    &-     &0.4   &0.2   &-     &0.3   &0.4   &-     &0.3   &0.8\\
\ S/H $^5$    &-      &0.1     &0.3    &-     &0.2   &0.1   &-     &0.02  &0.02  &-     &0.2   &0.3\\
\ str         &-      &0       &1      &-     &0     &1     &-     &0     &1     &-     &0     &1\\ \hline
\end{tabular}

$^0$ \erg; $^1$ \kms; $^2$ \cm3; $^3$ parsec; $^4$ photon cm$^{-2}$ s$^{-1}$ eV$^{-1}$ at the Lyman limit; $^5$ in 10$^{-4}$ units

\end{table*}

\begin{table*}
\centering
\caption{ Comparison of observed with calculated line ratios to \Hb=1 for the Nakajima et al. galaxies}
\begin{tabular}{llccccccccccccc} \hline  \hline
\   HSC      &N1    &ms1.0 &ms1.1&N3   &ms3.0&ms3.1& N5   &ms5.0&ms5.1**& N6  &ms6.0**&ms6.1\\ \hline
\ [OII]3727+ &0.42 &0.4    &0.45 &0.21  & 0.25 &0.24  & 0.43 &0.414&0.55&0.825 &0.84&0.86\\
\ [NeIII]3896+&0.29&0.24   &0.3  &0.25  &0.29  &0.27  & 0.71 &0.69 &0.70&0.42  &0.34 &0.37\\
\ \Hg 4340    &0.443&0.46  &0.44 &0.445 &0.46  &0.46  & 0.45 &0.46 &0.45&0.5   &0.46 &0.46\\       
\ [OIII]4363  &0.085&0.074 &0.06 &0.08  &0.1   &0.056 & 0.167&0.21 &0.12&0.124 &0.08 &0.09\\
\ HeII 4686   &$<$0.01&0.005&0.01 &0.024 &0.02  &0.05  &$<$0.013 &0.01 &0.008&0.03 &0.002*&0.024\\
\ \Hb 4861    &1    &1     &1    & 1    &1     &1     & 1    &1    &1   &1     &1    &1   \\
\ [OIII]5007+ &3.94 &3.72  &3.8  &3.25  & 3.42 &3.2   &7.2   &7.04 &7.7 &6.36  &6.5  &6.99\\
\ HeI 5876    &0.21 &0.08  &8e-5 &0.086 &0.16  &0.014 &0.113 &0.015&0.004&0.094&0.15 &1.25\\ 
\ \Ha 6563    &2.67 &2.9   &3.5  &2.53  &2.9   &3.3   & 2.84 &2.9  &2.9  &2.9  &2.89 &2.88\\
\ [NII]6583   &0.036&0.036 &0.05&0.007  &0.01  &0.009 &-     &0.01 &0.012&0.026&0.02  &0.03\\ 
\ [SII]6716   &0.04  &0.01 &0.01 &0.023 & 0.01 &0.01  &0.032 &0.01 &0.02 &0.089&0.02 &0.014 \\
\ [SII]6730   &0.02  &0.03 &0.02 &0.01  & 0.022&0.024 &0.03  &0.02 &0.05 &0.06 &0.05 &0.031\\
\ \Hb$_{calc}$ $^0$ &-&0.26&17.0 &-     &0.1   &3.0   &-     &0.09 &0.28 &-    &0.45 &0.064\\
\ \Vs $^1$    &-     &300  &280  &  -   & 300  &220   &-     &300  &300  &-    &180  &160\\
\ \n0 $^2$    &-     &1800 &600  &  -   & 2300 &3000  &-     &2000 &1900 &-    &1500 &1500\\
\  $D$ $^3$   &-     &2.9e-4&8.3 &-     &2.2e-4&0.007 &-     &1.7e-4&1.5e-4&-   &2.e-3&2.1e-4\\
\ \Ts $^4$    &-     &5.6  &13.0 & -    & 5.5  &11.0  &-     &6.8   &11. &-    &4.6   &3.9   \\
\ $U$         &      &0.07 &9.6  & -    & 0.06 &0.07  &-     &0.055 &0.042&-    &0.12  &0.21\\
\ He/H        &-     &0.01 &0.001&  -   & 0.1  &0.02  &-     &0.01&0.003 &-    &0.1   &0.8  \\
\ N/H $^5$    &-     &0.04 &0.05 &  -   & 0.01 &0.01  &-     &0.005&0.005&-    &0.01  &0.03 \\
\ O/H $^5$    &-     &5.3  &3.4  &  -   & 2.4  &5.0   &-     &5.7  &6.1  &-    &5.0   &3.6\\
\ Ne/H $^5$   &-     &0.3  &0.3  &  -   &  0.2 &0.3   &-     &0.44 &0.3  &-    &0.44  &0.44\\
\ S/H $^5$    &-     &0.3  &0.02 &  -   & 0.2  &0.03  &-     &0.2  &0.2  &-    &0.2   &0.2\\
\ str         &-     &0    &1    & -    &0     &1     &-     &0    &1    &-    &0     &1   \\ \hline
\end{tabular}

$^0$ \erg; $^1$ \kms; $^2$ \cm3; $^3$ parsec; $^4$ 10$^4$K; $^5$ in 10$^{-4}$ units; * the line can be blended with
	[FeIII]4702 with ratios to \Hb~ 0.06; ** : models calculated by \B0=6$\times$10$^{-5}$Gauss.

\end{table*}

\begin{table*}
\centering
\caption{ Comparison of observed with calculated line ratios to \Hb=1 for the Nakajima et al. galaxies}
\begin{tabular}{llccccccccccccc} \hline  \hline
        \    HSC      &N7     &ms7.0&ms7.1&N8    &ms8.0 &ms8.1 & N9    &ms9.0&ms9.1\\ \hline
        \ [OII]3727+  &0.82  &1.0   &0.83  &0.8  &0.82   &1.1    &0.17  &0.164 &0.18 \\
        \ [NeIII]3896+&0.34  &0.3   &0.4   &0.74 &0.78   &0.6    &0.23  &0.28  &0.35 \\
        \ \Hg 4340    &0.47  &0.46  &0.46  &0.53 &0.46   &0.45   &0.46  &0.46  &0.46 \\
        \ [OIII]4363  &$<$0.026&0.024&0.015&0.10 &0.126  &0.1    &0.078 &0.064 &0.055 \\
        \ HeII4686    &-     &0.002 &0.002 &-    &0.01   &0.02   &$<$0.013&0.007&0.015 \\
        \ \Hb 4861    &1     &1     &1     &1    &1      &1      &1       &1    &1    \\
        \ [OIII]5007+ &4.0   &4.4   &3.9   &10.0 & 9.9   &9.0    &3.33    &4.1  &3.5  \\
        \ HeI 5876    &0.092 &0.12  &0.12  &0.106&0.11   &0.003  &0.13    &0.026&0.0012\\
        \ \Ha 6563    &2.78  & 2.9  &2.9   &3.0  &2.9    &3.3    &2.83    &2.9  &3.11  \\
        \ [NII]6583   &-     &0.007 &0.01  &0.03 &0.035  &0.046   & $<$0.012&0.007&0.01 \\
        \ [SII]6716   &0.094 &0.013 &0.12  &-    &0.014  &0.02   &0.016    &0.016&0.013  \\
        \ [SII]6730   &0.1   &0.027 &0.25  &0.05 &0.031  &0.05   &0.013    &0.04 &0.031 \\
        \ \Hb$_{calc}$ $^0$ &-&3.2  &0.50  &-    &0.156  &1.34   &-        &2.28 &5.9   \\
        \ \Vs $^1$    &-    &300    &300   &-    &260    &290    &-        &250  &400  \\
        \ \n0 $^2$    &-    &1500   &1500  &-    &1500   & 600   & -       &2000 &2200 \\
	\  $D$ $^3$   &-    &0.03   &0.023 &-    &6.6e-4 &4.3e-3 &-        &3.2e-3&5.3e-3  \\
        \ \Ts $^4$    &-    &3.5    &4.2   &-    &5.3    &15.    &-        &6.9  &12    \\
	\ $U$         &-    &0.1    &0.24  &-    & 0.08  &0.19   &-        &0.15 &0.29   \\
        \ He/H        &-    &0.08   &0.08  &-    &0.07   &0.002  &-        &0.02    &0.002 \\
        \ N/H $^5$    &-    &0.01   &0.01  &-    &0.03   &0.02   &-        &0.01    &0.01  \\
        \ O/H $^5$    &-    &4.0    &4.0   &-    &5.3    &3.8    &-        &2.9     &3.0  \\
        \ Ne/H $^5$   &-    &1.0    &1.0   &-    &0.6    &0.2    &     -   &0.24    &0.18 \\
        \ S/H $^5$    &-    &0.2    &0.06  &-    &0.2    &0.01   &-        &0.2     &0.03 \\
        \ str         &-    &0      &1     &-    &0      &1      &-        &0       &1     \\ \hline
\end{tabular}

	$^0$ \erg; $^1$ \kms; $^2$ \cm3; $^3$ parsec; $^4$ 10$^4$K; $^5$ in 10$^{-4}$ units

\end{table*}

\begin{table*}
\centering
\caption{ Comparison of observed with calculated line ratios to \Hb=1 for the Nakajima et al. galaxies}
\begin{tabular}{llccccccccccccc} \hline  \hline
\  SDSS       & N10    &ms10.0 &ms10.1  &N11  & ms11.0& ms11.1&N12 &ms12.0&ms12.1&N13   &ms13.0&ms13.1 \\ \hline
\ [OII]3727+  &0.22   &0.19    &0.19   &0.85  &0.85  &1.10  &0.176 &0.23 &0.2      & 0.48  &0.47  & 0.46\\
\ [NeIII]3896+&0.42   &0.42    &0.42   &0.52  &0.45  &0.54  &0.73  &0.73 &0.62    & 0.517 &0.6   &0.49\\
\ \Hg 4340    &0.48   &0.46    &0.46   &0.49  &0.47  &0.44  &0.49  &0.46 &0.46    & 0.48  &0.47  &0.46\\
\ [OIII]4363  &0.134  &0.09    &0.11   &0.11  &0.11  &0.13  &0.22  &0.19 &0.21    & 0.149 &0.16  &0.14\\
\ HeII 4686   &0.018  &0.02    &0.02   &0.017 &0.014 &0.03  &0.008 &0.008&0.009   & 0.022 &0.02  &0.022\\
\ \Hb 4861    &1      &1       &1      &1     &1     &1     &1     &1    &1       & 1     &1     &1\\
\ [OIII]5007+ &5.8    &6.38    &5.7    &10.02 &9.0   &9.14  &10.41 &9.45 &9.9     & 7.29  &8.3   &7.7\\
\ HeI 5876    &0.086  &0.012   &0.002  &0.11  &0.41  &0.001 &0.097 &5.e-4&4.e-4   & 0.101 &0.14  &0.003\\
\ \Ha 6563    &2.65   &2.9     &3.0    &2.74  &2.87  &3.7   &2.88  &3.0  &3.1     & 2.84  &2.9   &3.2\\
\ [NII]6583   &0.008  &0.007   &0.008  &0.15  &0.2   &0.14  &0.008 &0.013&0.01    &0.017 &0.017 &0.018\\
\ [SII]6716   &0.02   &0.016   &0.008  &0.072 &0.03  &0.02  &-     &8e-3 &3e-3    &0.046 &0.014 &0.03\\
\ [SII]6730   &0.016  &0.037   &0.019  &0.077 &0.07  &0.06  &-     &0.02  &7.5e-3 &0.031 &0.31  &0.06\\
\ \Hb$_{calc}$$^0$&-  &2.3     &6.2    &-     &0.13  &1.7   &-     &0.72  &7.1    &-     &0.18  &4.0 \\
\ \Vs $^1$    &-     &320      &440    &-     &300   &300   &-     &320   &260    &-     &300   &380\\
\ \n0 $^2$    &-      &2000    &2300   &-     &1500  & 900  &-     &2000  &3200   &-     &1500  &2600 \\
\  $D$ $^3$   & -     &0.002   &3.3e-4 &-     &1.1e-3&3.3e-3&-     &8.3e-4&3.3e-4 &-     &6.e-4 &1.27e-3\\
\ \Ts $^4$    &-      &8.5     &12      &-     &4.5   & 18.0 &-     &15.0  &10.0   &-     &5.0   &15.0\\
\ $U$         &-      &0.13    &0.27   &-     &0.15  &0.3   &-     & 0.099 &0.6   &-     &0.1  &0.16\\
\ He/H        &-      &0.01    &0.003  &-     &0.26  &0.003 &-     &0.001 &0.001  &-     &0.09 &0.0026\\
\ N/H $^5$    &-      &0.01    &0.01   &-     &0.16  &0.04  &-     &0.01  &0.015  &-     &0.019  &0.014\\
\ O/H $^5$    &-      &4.0     &3.8    &-     &4.2   &3.0   &-     &2.3   &4.0    &-     &3.5   &6.60\\
\ Ne/H $^5$   &-      &0.3     &0.2    &-     &0.4   &0.16  &-     &0.2   &0.22   &-     &0.4   &0.2\\
\ S/H $^5$    &-      &0.2     &0.03   &-     &0.5   &0.01  &-     &0.1   &0.01   &-     &0.3   &0.05\\
\ str         &-      &0       &1      &-     &0     &1     &-     &0     &1      &-     &0     &1\\ \hline
\end{tabular}

 $^0$ \erg; $^1$ \kms; $^2$ \cm3; $^3$ parsec; $^4$ 10$^4$K; $^5$ in 10$^{-4}$ units

\end{table*}

\end{appendix}

\end{document}